\documentclass[iop,revtex4-1]{emulateapj}
\usepackage{aas_macros}
\usepackage[dvipsnames,svgnames]{xcolor}
\usepackage{color,soul}
\usepackage{natbib}
\usepackage{hyperref}
\usepackage{amsmath}
\usepackage{xspace}

\usepackage{graphicx}
\usepackage{enumitem}
\usepackage{amssymb}
\usepackage{xifthen}
\usepackage{hyperref}

\hypersetup{    
  colorlinks      = {true},
  linkcolor       = {blue},
  citecolor       = {blue},
  urlcolor        = {blue},
}

\newcommand{\code}[1]{\texttt{#1}\xspace}

\newcommand{\unit}[1]{\ensuremath{\mathrm{\,#1}}\xspace}

\newcommand*\ruleline[1]{\par\noindent\raisebox{1ex}{\makebox[0.97\linewidth]{\hrulefill\quad\raisebox{-.2ex}{#1}\quad\hrulefill}}}

\newcommand{\feh}         {{\rm [Fe/H]}}
\newcommand{\vhel}         {\mbox{$v_{\mathrm{hel}}$}}
\newcommand{\sigmav}         {\mbox{${\sigma_v}$}}
\newcommand{\sigmafeh}         {\mbox{$\sigma_{\mathrm{[Fe/H]}}$}}

\newcommand{\Beta}{B\xspace}

\newcommand{\km}{\unit{km}}

\newcommand{\kpc}{\unit{kpc}}

\newcommand{\second}{\unit{s}}
\newcommand{\kms}         {\ensuremath{\km\,\second^{-1}}\xspace}

\newcommand{\msun}{\unit{M_\odot}}
\newcommand{\lsun}{\unit{L_\odot}}

\newcommand{\magn}{\unit{mag}}

\newcommand{\vbulk}{\ensuremath{-101.2 \pm 0.5}\xspace}

\newcommand{\vdisp}{\ensuremath{0.9^{+0.6}_{-0.5} }\xspace}
\newcommand{\vgrad}{\ensuremath{-8.0\pm0.4 }\xspace}
\newcommand{\PA}{\ensuremath{81\degr \pm 14\degr}\xspace}
\newcommand{\stellarmass}{\ensuremath{3.8\times10^3 }\xspace}

\newcommand{\fehmean}{\ensuremath{-2.49 \pm 0.06}\xspace}
\newcommand{\fehdisp}{\ensuremath{0.11 ^{+0.07}_{-0.06}}\xspace}

\newcommand{\gaia}{\textit{Gaia}\xspace}

\newcommand{\noinfo}{......\xspace}

\shorttitle{Spectroscopic Analysis of Tucana III Stream}
\shortauthors{Li et~al.}

\begin{document}

\title{The First Tidally Disrupted Ultra-Faint Dwarf Galaxy? -- Spectroscopic Analysis of the Tucana III Stream\altaffilmark{*}\altaffilmark{\dag}}

\altaffiltext{*}{This paper includes data gathered with Anglo-Australian Telescope in Australia.}
\altaffiltext{\dag}{This paper includes data gathered with the 6.5 meter
  Magellan Telescopes located at Las Campanas Observatory, Chile.}


\def\andname{}
\author{
T.~S.~Li\altaffilmark{1,2},
J.~D.~Simon\altaffilmark{3},
K.~Kuehn\altaffilmark{4},
A.~B.~Pace\altaffilmark{5},
D.~Erkal\altaffilmark{6,7},
K.~Bechtol\altaffilmark{8},
B.~Yanny\altaffilmark{1},
A.~Drlica-Wagner\altaffilmark{1},
J.~L.~Marshall\altaffilmark{5},
C.~Lidman\altaffilmark{4,9},
E.~Balbinot\altaffilmark{6},
D.~Carollo\altaffilmark{10,11},
S.~Jenkins\altaffilmark{2},
C.~E.~Mart{\'\i}nez-V{\'a}zquez\altaffilmark{12},
N.~Shipp\altaffilmark{2},
K.~M.~Stringer\altaffilmark{5},
A.~K.~Vivas\altaffilmark{12},
A.~R.~Walker\altaffilmark{12},
R.~H.~Wechsler\altaffilmark{13,14,15},
F.~B.~Abdalla\altaffilmark{16,17},
S.~Allam\altaffilmark{1},
J.~Annis\altaffilmark{1},
S.~Avila\altaffilmark{18},
E.~Bertin\altaffilmark{19,20},
D.~Brooks\altaffilmark{16},
E.~Buckley-Geer\altaffilmark{1},
D.~L.~Burke\altaffilmark{14,15},
A.~Carnero~Rosell\altaffilmark{21,22},
M.~Carrasco~Kind\altaffilmark{23,24},
J.~Carretero\altaffilmark{25},
C.~E.~Cunha\altaffilmark{14},
C.~B.~D'Andrea\altaffilmark{26},
L.~N.~da Costa\altaffilmark{21,22},
C.~Davis\altaffilmark{14},
J.~De~Vicente\altaffilmark{27},
P.~Doel\altaffilmark{16},
T.~F.~Eifler\altaffilmark{28,29},
A.~E.~Evrard\altaffilmark{30,31},
B.~Flaugher\altaffilmark{1},
J.~Frieman\altaffilmark{1,2},
J.~Garc\'ia-Bellido\altaffilmark{32},
E.~Gaztanaga\altaffilmark{33,34},
D.~W.~Gerdes\altaffilmark{30,31},
D.~Gruen\altaffilmark{14,15},
R.~A.~Gruendl\altaffilmark{23,24},
J.~Gschwend\altaffilmark{21,22},
G.~Gutierrez\altaffilmark{1},
W.~G.~Hartley\altaffilmark{16,35},
D.~L.~Hollowood\altaffilmark{36},
K.~Honscheid\altaffilmark{37,38},
D.~J.~James\altaffilmark{39},
E.~Krause\altaffilmark{28,29},
M.~A.~G.~Maia\altaffilmark{21,22},
M.~March\altaffilmark{26},
F.~Menanteau\altaffilmark{23,24},
R.~Miquel\altaffilmark{40,25},
A.~A.~Plazas\altaffilmark{29},
E.~Sanchez\altaffilmark{27},
B.~Santiago\altaffilmark{41,21},
V.~Scarpine\altaffilmark{1},
R.~Schindler\altaffilmark{15},
M.~Schubnell\altaffilmark{31},
I.~Sevilla-Noarbe\altaffilmark{27},
M.~Smith\altaffilmark{42},
R.~C.~Smith\altaffilmark{12},
M.~Soares-Santos\altaffilmark{43},
F.~Sobreira\altaffilmark{44,21},
E.~Suchyta\altaffilmark{45},
M.~E.~C.~Swanson\altaffilmark{24},
G.~Tarle\altaffilmark{31},
D.~L.~Tucker\altaffilmark{1}
\\ \vspace{0.2cm} (DES Collaboration) \\
}
\affil{$^{1}$ Fermi National Accelerator Laboratory, P. O. Box 500, Batavia, IL 60510, USA}\email{tingli@fnal.gov}
\affil{$^{2}$ Kavli Institute for Cosmological Physics, University of Chicago, Chicago, IL 60637, USA}
\affil{$^{3}$ Observatories of the Carnegie Institution for Science, 813 Santa Barbara St., Pasadena, CA 91101, USA}
\affil{$^{4}$ Australian Astronomical Observatory, North Ryde, NSW 2113, Australia}
\affil{$^{5}$ George P. and Cynthia Woods Mitchell Institute for Fundamental Physics and Astronomy, and Department of Physics and Astronomy, Texas A\&M University, College Station, TX 77843,  USA}
\affil{$^{6}$ Department of Physics, University of Surrey, Guildford GU2 7XH, UK}
\affil{$^{7}$ Institute of Astronomy, University of Cambridge, Madingley Road, Cambridge CB3 0HA, UK}
\affil{$^{8}$ LSST, 933 North Cherry Avenue, Tucson, AZ 85721, USA}
\affil{$^{9}$ The Research School of Astronomy and Astrophysics, Australian National University, ACT 2601, Australia}
\affil{$^{10}$ INAF Osservatorio Astrofisico di Torino, Via Osservatorio 20, 10025 Pino Torinese, Italy}
\affil{$^{11}$ ARC Centre of Excellence for All-sky Astrophysics (CAASTRO)}
\affil{$^{12}$ Cerro Tololo Inter-American Observatory, National Optical Astronomy Observatory, Casilla 603, La Serena, Chile}
\affil{$^{13}$ Department of Physics, Stanford University, 382 Via Pueblo Mall, Stanford, CA 94305, USA}
\affil{$^{14}$ Kavli Institute for Particle Astrophysics \& Cosmology, P. O. Box 2450, Stanford University, Stanford, CA 94305, USA}
\affil{$^{15}$ SLAC National Accelerator Laboratory, Menlo Park, CA 94025, USA}
\affil{$^{16}$ Department of Physics \& Astronomy, University College London, Gower Street, London, WC1E 6BT, UK}
\affil{$^{17}$ Department of Physics and Electronics, Rhodes University, PO Box 94, Grahamstown, 6140, South Africa}
\affil{$^{18}$ Institute of Cosmology \& Gravitation, University of Portsmouth, Portsmouth, PO1 3FX, UK}
\affil{$^{19}$ CNRS, UMR 7095, Institut d'Astrophysique de Paris, F-75014, Paris, France}
\affil{$^{20}$ Sorbonne Universit\'es, UPMC Univ Paris 06, UMR 7095, Institut d'Astrophysique de Paris, F-75014, Paris, France}
\affil{$^{21}$ Laborat\'orio Interinstitucional de e-Astronomia - LIneA, Rua Gal. Jos\'e Cristino 77, Rio de Janeiro, RJ - 20921-400, Brazil}
\affil{$^{22}$ Observat\'orio Nacional, Rua Gal. Jos\'e Cristino 77, Rio de Janeiro, RJ - 20921-400, Brazil}
\affil{$^{23}$ Department of Astronomy, University of Illinois at Urbana-Champaign, 1002 W. Green Street, Urbana, IL 61801, USA}
\affil{$^{24}$ National Center for Supercomputing Applications, 1205 West Clark St., Urbana, IL 61801, USA}
\affil{$^{25}$ Institut de F\'{\i}sica d'Altes Energies (IFAE), The Barcelona Institute of Science and Technology, Campus UAB, 08193 Bellaterra (Barcelona) Spain}
\affil{$^{26}$ Department of Physics and Astronomy, University of Pennsylvania, Philadelphia, PA 19104, USA}
\affil{$^{27}$ Centro de Investigaciones Energ\'eticas, Medioambientales y Tecnol\'ogicas (CIEMAT), Madrid, Spain}
\affil{$^{28}$ Department of Astronomy/Steward Observatory, 933 North Cherry Avenue, Tucson, AZ 85721-0065, USA}
\affil{$^{29}$ Jet Propulsion Laboratory, California Institute of Technology, 4800 Oak Grove Dr., Pasadena, CA 91109, USA}
\affil{$^{30}$ Department of Astronomy, University of Michigan, Ann Arbor, MI 48109, USA}
\affil{$^{31}$ Department of Physics, University of Michigan, Ann Arbor, MI 48109, USA}
\affil{$^{32}$ Instituto de Fisica Teorica UAM/CSIC, Universidad Autonoma de Madrid, 28049 Madrid, Spain}
\affil{$^{33}$ Institut d'Estudis Espacials de Catalunya (IEEC), 08193 Barcelona, Spain}
\affil{$^{34}$ Institute of Space Sciences (ICE, CSIC),  Campus UAB, Carrer de Can Magrans, s/n,  08193 Barcelona, Spain}
\affil{$^{35}$ Department of Physics, ETH Zurich, Wolfgang-Pauli-Strasse 16, CH-8093 Zurich, Switzerland}
\affil{$^{36}$ Santa Cruz Institute for Particle Physics, Santa Cruz, CA 95064, USA}
\affil{$^{37}$ Center for Cosmology and Astro-Particle Physics, The Ohio State University, Columbus, OH 43210, USA}
\affil{$^{38}$ Department of Physics, The Ohio State University, Columbus, OH 43210, USA}
\affil{$^{39}$ Harvard-Smithsonian Center for Astrophysics, Cambridge, MA 02138, USA}
\affil{$^{40}$ Instituci\'o Catalana de Recerca i Estudis Avan\c{c}ats, E-08010 Barcelona, Spain}
\affil{$^{41}$ Instituto de F\'\i sica, UFRGS, Caixa Postal 15051, Porto Alegre, RS - 91501-970, Brazil}
\affil{$^{42}$ School of Physics and Astronomy, University of Southampton,  Southampton, SO17 1BJ, UK}
\affil{$^{43}$ Brandeis University, Physics Department, 415 South Street, Waltham MA 02453}
\affil{$^{44}$ Instituto de F\'isica Gleb Wataghin, Universidade Estadual de Campinas, 13083-859, Campinas, SP, Brazil}
\affil{$^{45}$ Computer Science and Mathematics Division, Oak Ridge National Laboratory, Oak Ridge, TN 37831}


\begin{abstract}
We present a spectroscopic study of the tidal tails and core of the Milky Way satellite Tucana~III, collectively referred to as the Tucana~III stream, using the 2dF+AAOmega spectrograph on the Anglo-Australian Telescope and the IMACS spectrograph on the Magellan/Baade Telescope. In addition to recovering the brightest 9 previously known member stars in the Tucana~III core, we identify 22 members in the tidal tails. We observe strong evidence for a velocity gradient of $8.0\pm0.4~\mathrm{\km\,\second^{-1}}\,\mathrm{deg}^{-1}$ over at least 3\degr\ on the sky. Based on the continuity in velocity we confirm that the Tucana~III tails are real tidal extensions of Tucana~III. The large velocity gradient of the stream implies that Tucana~III is likely on a radial orbit. We successfully obtain metallicities for 4 members in the core and 12 members in the tails. We find that members close to the ends of the stream tend to be more metal-poor than members in the core, indicating a possible metallicity gradient between the center of the progenitor halo and its edge. The spread in metallicity suggests that the progenitor of the Tucana~III stream is likely a dwarf galaxy rather than a star cluster. Furthermore, we find that with the precise photometry of the Dark Energy Survey data, there is a discernible color offset between metal-rich disk stars and metal-poor stream members. This metallicity-dependent color offers a more efficient method to recognize metal-poor targets and will increase the selection efficiency of stream members for future spectroscopic follow-up programs on stellar streams.
\end{abstract}


\section{INTRODUCTION}
\label{intro}


Stellar streams, originating from the tidal disruption of dwarf galaxies~\citep[e.g., the Sagittarius stream; ][]{Majewski2003, Belokurov2006b} and globular clusters~\citep[e.g., the Palomar 5 tidal tails;][]{Odenkirchen2001}, are excellent tracers for probing the underlying shape of the Milky Way's dark matter halo 
\citep[see, e.g.,][]{Johnston:2005,Law:2010,Bovy:2014,Bonaca:2014,Kupper2015ApJ...803...80K,Erkal:2016,Bovy:2016}.
Density perturbations along kinematically cold (i.e., small velocity dispersion) stellar streams can additionally be used to measure the abundance of low-mass dark matter substructure \citep[e.g.,][]{Ibata:2002,Johnston:2002,Carlberg:2009,yoon_etal_2011,Erkal:2015,Carlberg:2016a,Bovy:2017,Erkal:2017}
and therefore provide a direct test of the standard $\Lambda$CDM cosmological model~\citep{Springel:2008}. These diffuse stellar features also inform us that our Galactic stellar halo is shaped by the merging of neighboring smaller galaxies, which is predicted by hierarchical structure formation models of galaxy evolution~\citep{Peebles:1965, Searle1978, Steinmetz2002, Bullock2005, Font2011}.  We refer readers to~\citet{Newberg:2016} for a recent review.

Thanks to the unprecedented photometric precision, depth, and coverage area of the Dark Energy Survey~\citep[DES;][]{DES:2005,DES:2016,DES:2017,desdr1}, more than a dozen new stellar streams have been discovered using the first three years of data ~\citep{dw15b,Balbinot:2016,shipp2018}. Of the streams discovered in DES, the Tucana~III (Tuc~III) stellar stream is among the most intriguing because it is the only new stream that has an unambiguous progenitor identified. 

The Tuc~III stellar stream was discovered as a pair of very low surface brightness linear features adjacent to the ultra-faint dwarf galaxy candidate Tuc~III \citep{dw15b}. Located at a heliocentric distance of just $\sim25$ kpc and a Galactocentric distance of $\sim23$ kpc, the stream extends at least 2 degrees from either side of Tuc~III.  Further analyses conducted by~\citet{shipp2018} showed that the stream has a projected length of 2 kpc (4.8\degr) and a width of 79 pc ($\sigma_w = 0.18\degr$) on the sky. 
The stream is the result of the on-going disruption of Tuc~III. Following the terminology that was used in~\citet{shipp2018}, in this paper we will refer to the Tuc~III dwarf galaxy (candidate) as the Tuc~III core, the tidal tails of Tuc~III as the Tuc~III tails, and the whole system as the Tuc~III stream. 
{Specifically, we consider all Tuc~III member stars confirmed in~\citet{tuc3} as belonging to the Tuc~III core. We define the radius of Tuc~III core $r$ to be the radial distance to the outermost Tuc~III member confirmed in~\citet{tuc3}, i.e. a radius of $r \lesssim 0.22\degr$ (or $\sim97$ pc), or 2.2 times the half-light radius of the Tuc~III dwarf galaxy candidate determined in~\citet{dw15b}. }

If Tuc~III is a dwarf galaxy, the Tuc~III stream will be a prototype for the tidal disruption of the smallest galaxies. Spectroscopic observations were conducted in the Tuc~III core by~\citet{tuc3} with Magellan/IMACS, but the authors were not able to conclusively classify Tuc~III as an ultra-faint dwarf galaxy or star cluster due to the unresolved velocity and metallicity dispersions. \citet{tuc3} tentatively suggested that Tuc~III is the tidally-stripped remnant of a dark matter-dominated dwarf galaxy, based on its large size and low mean metallicity. 

{Though dozens of streams have been found in the Milky Way halo, only a handful of streams have an unambiguous progenitor identified~\citep[e.g. the Sagittarius stream, the Palomar 5 tidal tails, and NGC 5466 tidal tails][ ]{Belokurov2006c}. }
The Tuc~III stream offers an opportunity to investigate mechanisms of tidal disruption and resulting formation of tidal streams in great detail.  
Furthermore, the presence of the progenitor also makes the stream an ideal candidate for orbit fitting, as well as a valuable tracer of the Milky Way's gravitational potential.

In this paper, we present results from spectroscopic observations of the Tuc~III stream using the 2dF+AAOmega spectrograph on the Anglo-Australian Telescope and the IMACS spectrograph on the Magellan/Baade Telescope. We describe the observations and data reduction from these two instruments in \S\ref{sec:observations}. We identify member stars of the stream and determine the kinematic and metallicity properties of the stream in \S\ref{sec:results}. In \S\ref{sec:discuss}, we discuss the properties of the stream and the comparisons with other known streams and dwarf galaxies. In \S\ref{sec:colorcolor} we demonstrate the use of photometric measurements to select candidate metal-poor stream members based on their colors for future spectroscopic follow-up programs. We conclude in \S\ref{sec:summary}. 

\section{OBSERVATIONS AND DATA REDUCTION}
\label{sec:observations}

\subsection{AAT/2df+AAOmega Observations}\label{sec:aat}

We observed candidate member stars in the Tuc~III stream with the AAOmega Spectrograph~\citep{Sharp2006} on the 3.9~m Anglo-Australian Telescope (AAT).  AAOmega is a dual-beam spectrograph, which feeds a blue arm and a red arm with a beam splitter at 5700~\AA. This paper focuses on the spectra obtained with the red arm using the 1700D grating, which has a spectral resolution of $R = 10000$, a pixel scale of 0.23~\AA/pixel, and a wavelength coverage of $8400-8810$~\AA. This wavelength range contains the Calcium triplet (CaT) absorption lines which are the primary source of velocity and metallicity measurements.  

The AAOmega Spectrograph is fed by the Two Degree Field (``2dF") fiber positioner facility, allowing the acquisition of up to 400 simultaneous spectra of objects within a 2\degr\ field in diameter on the sky. Among the 400 fibers, 25 are assigned to sky positions and 8 are assigned to guide stars selected from the UCAC4 catalog~\citep{Zacharias:2013}.  The remaining fibers are assigned to the target stars.
 
We selected the targets using an empirical color-magnitude locus derived from the confirmed member stars in the Tuc~III dwarf galaxy~\citep{tuc3}. Based on the location of candidate stars on the sky and on the color-magnitude diagram using an early version of DES astrometry and photometry, namely the Y2Q catalog~\citep{dw15b}, a membership probability for each star was calculated in a similar way as discussed in~\citet{dw15b}. We then prioritize targets based on a combination of brightness and membership probability, and allocated targets to the fibers using the $configure$\footnote{$configure$ provides a graphic user interface for fiber allocations, see details at \url{https://www.aao.gov.au/science/software/configure}} software provided by Australian Astronomical Observatory. As shown in Figure~\ref{fig:targets}, we selected several hundred main-sequence turnoff (MSTO) stars and red giant branch (RGB) stars, a few dozen red horizontal branch (RHB) stars and a handful of blue horizontal branch (BHB) stars.

Observations were conducted on 2016 June 30, July 9--13, July 25--27, and 2017 August 22. We had a total of 9 half nights of observing time in 2016 and 2 hours of service time in 2017. The exposures are typically composed of several 30-40 min exposures. 
Due to unfavorable weather conditions, we did not observe on July 11. The dome was partially closed for many nights due to clouds, including June 30, July 9, July 13, July 26 and July 27. Exposures on July 12 and 13 were taken through thick clouds and were therefore excluded from the analysis. About half of the exposures taken on July 10 were also discarded due to the presence of thin clouds.  Specifically, data are excluded from the analysis when the extracted 1D spectra from one exposure have signal-to-noise ratios (S/N) of less than 2~pixel$^{-1}$ for stars at $g\sim18$. As a comparison, with a 30 min exposure, the S/N is around 7~pixel$^{-1}$ for stars at $g\sim18$ in the absence of clouds. Overall, we lost roughly 6 of the 9 half-nights due to poor weather. The remaining useful 3 half-nights have an average seeing of $\sim2.5\arcsec$.

Since the length of the stream was determined to be about 4\degr\ at the time of discovery~\citep{dw15b}, we targeted the Tuc~III stream with two telescope pointings whose centers are offset by 0.8\degr\ west (\code{AAT-Field-1}) and east (\code{AAT-Field-2}) from the center of Tuc~III core during the 2016 classical observing time. This ensures that the Tuc~III core will get longer exposures in total and may reveal more faint members. Subsequent reanalysis by~\citet{shipp2018} revealed that the stream is slightly longer (4.8\degr) and we therefore added a third pointing (\code{AAT-Field-3}) 1.2\degr\ west from the center of Tuc~III for 2~hrs of observing on 2017 August 22 using service time. Due to the shorter exposure time, \code{AAT-Field-3} is shallower than the other two pointings. An illustration of the 3 fields on the sky along with the targets observed is shown in the top left panel of Figure~\ref{fig:targets}. We obtained S/N$\sim$15 per pixel for stars at $g\sim18$ in \code{AAT-Field-1} and \code{AAT-Field-2}; and S/N$\sim$9 per pixel for stars at $g\sim18$ in \code{AAT-Field-3}.  
{Though 2df has a total of 400 fibers (including 25 for sky fibers and 8 for guide stars) and a fiber collision radius of 30\arcsec-40\arcsec, flexibility in fiber allocation with 2dF allows us to change the targets from night to night and observe more targets at each field. Specifically, we performed quick data reduction and analysis on the observed spectra at the end of each night, measured the radial velocities of bright stars whose spectra have sufficient S/N (S/N $> 10$), and classified those stars that have velocity differences more than 100~\kms from the Tuc~III core velocity as non-members. We then re-allocated the fibers for those non-members to alternate targets in the subsequent night's observing.}
A total of 1045 candidate stars were observed over the entire program with AAT in three fields. 
{We observed roughly 85\% of the RGB candidates at $g < 19.5$ in the fields of 3 AAT pointings; the other 15\% candidates were not observed either due to fiber collision or due to low membership probability of the targets and the limited number of available fibers. We expect that the completeness of the true members at $g < 19.5$ is higher than 85\%, because the stars that are closer to the empirical color-magnitude locus will have a higher membership probability and therefore have a higher priority to be assigned to a fiber.}

The data reduction was performed using version 6.46 of  $2dfdr$\footnote{\url{https://www.aao.gov.au/science/software/2dfdr}}. The reduction includes bias subtraction, 2D scattered light subtraction, flat-fielding, Gaussian weighted spectral extraction, wavelength calibration, and sky subtraction. Wavelength calibration was first performed using the arc frames taken immediately before or after each science exposure, followed by a recalibration with a second order polynomial fit using sky emission lines. As the observations were taken from different nights, the reduced and extracted spectra were first corrected for the heliocentric motion of the Sun at each exposure. Then the spectra from multiple exposures were combined using inverse-variance weighting.  The final velocities and metallicities reported (see \S\ref{sec:measure}) are derived from the combined spectra over the entire program.

\subsection{Magellan/IMACS Observations}\label{sec:imacs}

\begin{figure*}[th!]
\centering
\epsscale{1}
\includegraphics[scale=0.45]{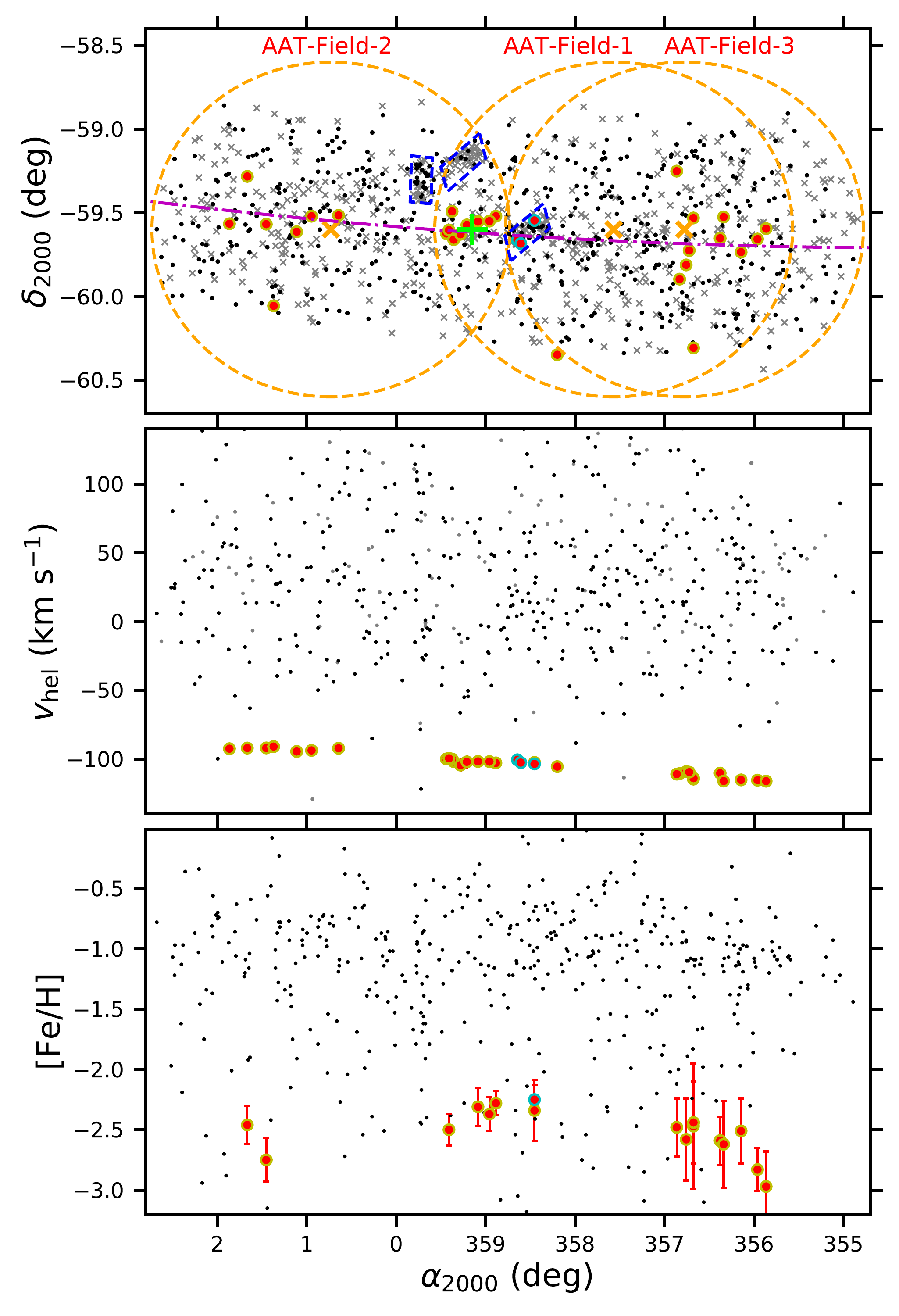}
\includegraphics[scale=0.42]{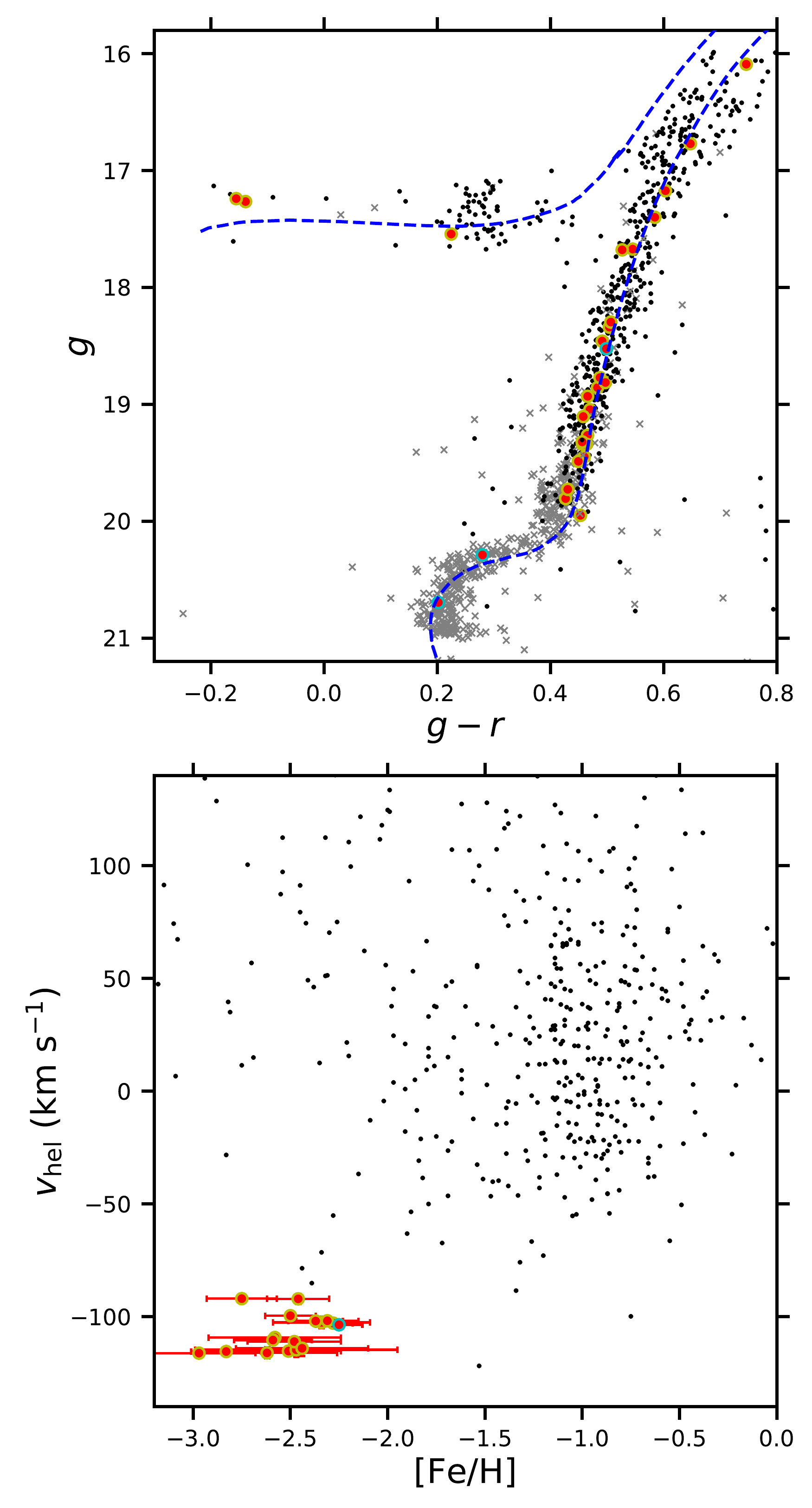}
\caption{\textbf{Top left:} Spatial distribution of the target stars in celestial coordinates ($\alpha_{2000}, \delta_{2000}$). The center of the Tuc~III dwarf galaxy is indicated with a lime plus symbol. 
The three orange ``x'' symbols indicate the center of the 3 AAT pointings along with the 3 orange circles to illustrate the field-of-view of AAT. The blue dashed boxes show the locations of 3 IMACS pointings.
The magenta dashed line shows a great circle on the sky using the end points of Tuc~III stream from~\citet{shipp2018}. The gray crosses represent the targets for which radial velocity measurements failed, while the black dots are the stars whose spectra have high enough S/N (S/N $\gtrsim$ 6) for radial velocity measurements. The red filled circles are the confirmed spectroscopic members of the Tuc~III stream from this work, with AAT members encircled in yellow and IMACS members encircled in cyan. (Symbols are the same for other panels and other figures if not specified.) 
\textbf{Top right:} Location of target stars in the color-magnitude diagram. The blue lines {show a Dotter isochrone~\citep{Dotter2008} for MSTO+RGB stars and a PARSEC isochrone~\citep{Bressan2012} for horizontal branch stars, both with} $\feh = -2.2$ and age = 12.5 Gyr at distance modulus $m-M = 17.0$.
\textbf{Middle left:} Measured heliocentric radial velocity of all targets. 31 member stars of the Tuc~III stream are grouped around $-100~\kms$. A clear trend of increasing velocity towards equatorial east (or larger $\alpha$) is evident from the spectroscopically confirmed members. The uncertainties on the velocities are smaller than the symbol sizes of the member stars. {Black (Gray) dots represents the stars for which the metallicity measurements are successful (unsuccessful) and are (not) shown in the bottom panel. }
\textbf{Bottom left:} A subset of the spectra have measured metallicities derived from the EWs of CaT lines (assuming these stars are at the distance of Tuc~III). Metallicities of 16 member stars in Tuc~III stream are successfully obtained. The metallicities of the members at the edges of the stream seem to be more metal-poor than the other members in the core of the stream. \textbf{Bottom right:} Scatter plot of the measured heliocentric velocity and metallicity of all the targets. The confirmed members are grouped around \feh\, $\sim -2.5$ and $v_\mathrm{hel}\sim-100~\kms$. Note that for better display, the error bars in \feh\, and $v_\mathrm{hel}$ are only shown for member stars. The error bars for the target stars are at a similar level as the member stars. 
}
\label{fig:targets}
\end{figure*}

In order to better probe the transition region ($0.3\degr \lesssim r \lesssim 0.5\degr$) between the (presumably) bound core of Tuc~III and the tidal tails, we also obtained additional spectroscopy of Tuc~III with the IMACS spectrograph \citep{dressler06} on the Magellan/Baade telescope.  We observed 3 slit masks on the nights of 2017 June 19 and 21.  Target selection for these masks followed the criteria described by \citet{tuc3}, 
and the {mask positions were chosen based on the highest densities of} bright candidate RGB stars.  
The spectrograph configuration was identical to those used 
by \citet{tuc3} and \citet{eri2}, with the f/4 camera and the 1200/32\fdg7 grating providing a spectral resolution of $R=11000$ over the wavelength range $\sim7500-8800$~\AA.  Observing conditions on June 19 for the first two masks were good (clear skies and seeing $<0\farcs8$), while the third mask on June 21 suffered from quite poor conditions (clouds and seeing $>1\arcsec$). 
{Two of the masks, which were both offset from the stream track, did not reveal any additional member stars, while the mask on the stream track identified two new members (both much fainter than what AAT can detect; see top right panel of Figure~\ref{fig:targets}).  These results suggest that Tuc~III members at these radii are concentrated in the unbound tails, and relatively few bright members of the core remain to be found.} 

The IMACS data were reduced as described by \citet{tuc3} and \citet{eri2}, employing a combination of the Cosmos reduction pipeline \citep{dressler11,oemler17} and a version of the DEEP2 data reduction pipeline \citep{cooper12,newman13} adapted for IMACS.

\subsection{Velocity and Metallicity Measurements}\label{sec:measure}

The reduced 1D spectra from AAT/2df+AAOmega (hereafter AAT) and from Magellan/IMACS (hereafter IMACS) were then used for radial velocity measurements following the same method as described in \citet{eri2, Li2018} and metallicity measurements following \citet{simon15b} and \citet{eri2}. We refer readers to these reference for more details and we briefly summarize the procedures below.

The radial velocities (RVs) were measured via template fitting using a set of radial velocity standards {(with various metallicities, temperatures and surface gravities)} observed with the same instrument setup and a maximum likelihood approach with a Markov chain Monte Carlo (MCMC) sampler. The statistical velocity uncertainties were determined from the standard deviation of the posterior velocity distribution from the MCMC sampler. We adopted a systematic floor of 0.5~\kms for AAT velocities~\citep{Li2018} and 1.0~\kms for IMACS velocities~\citep{tuc3,eri2} and added these systematic uncertainties in quadrature with the statistical uncertainties for each star to obtain the final reported velocity uncertainties. 

We also determined the metallicities of RGB candidate stars using the equivalent widths (EWs) of the CaT lines. We fit all three of the CaT lines with a Gaussian plus Lorentzian function and then converted the summed EWs of the three CaT lines to metallicity using the calibration relation as a function of absolute V magnitude from~\citet{carrera13}. We assume that the candidate stars are members of Tuc~III stream and are therefore at a distance of $25\pm2$~\kpc~\citep{dw15b} to derive the absolute magnitude of each candidate star. The uncertainties on the EWs are calculated from the Gaussian and Lorentzian fit plus a systematic uncertainty of {0.2~\AA\ \citep{eri2,Li2018} added in quadrature}.  The metallicity uncertainties are dominated by the uncertainties on the CaT EWs, with small contributions from the uncertainties on the distances, the stellar photometry, and the uncertainties on the calibration parameters from~\citet{carrera13}. We note that~\citet{shipp2018} find the Tuc~III stream spans 8~\kpc in distance. A 4~\kpc shift from 25~\kpc will cause roughly a 0.05 dex shift in \feh and is small compared to the uncertainties {from the EW measurement (typically at 0.25 dex)}.  We therefore excluded the possible distance gradient when computing the metallicity of member stars in the Tuc~III stream.

We applied the methods described above to the entire spectroscopic sample and report the derived RVs and metallicities in Table~\ref{tab:tuc3_spec}. We note that not all spectra have high enough S/N for adequate RV and EW fits. We assess the fitting quality of every spectrum visually. Usually, spectra that have S/N $<$ 6 do not provide a good RV fit and spectra that have S/N $<$ 9 do not provide a good EW fit. In total, we successfully determined the RVs of 552 candidate stars and EWs of 431 candidate stars from AAT. For IMACS, we determined the RVs of 57 candidates and EWs of 35 candidates. {In total, we have 13 spectra with SNR $>$ 6 were rejected because of a poor RV fit, while 29 spectra with SNR $>$ 9 were rejected because of a poor EW fit.}
We note that derived metallicities are only valid for stars that are truly RGB members of the Tuc~III stream. For the non-members, they are likely foreground main-sequence stars from the Milky Way at a different distance, so the calibration relation from~\citet{carrera13} does not apply to these stars. 

We note that even though we used the DES Y2Q catalog for target selection, the reported astrometry and photometry in Table~\ref{tab:tuc3_spec} are from a newer version, namely the DES DR1 catalog~\citep{desdr1}, and are used for analysis in later sections. Specifically, the weighted average magnitudes (\code{WAVG\_MAG\_PSF}) from DR1 are used throughout this work. The improved photometry precision in DES DR1 is especially important for the later discussion in~\S\ref{sec:colorcolor}. However, we noticed that $\sim2\%$ of the candidate stars present in the DES Y2Q catalog are missing in DES DR1.  This is due to an overly conservative rejection of stars that lie in the wings of nearby saturated stars when the weighted average magnitude (\code{WAVG\_MAG\_PSF}) quantities in DR1 are computed. For those stars, we present their photometry from the Y2Q catalog and mark them in Table~\ref{tab:tuc3_spec}. We note that the magnitudes reported here (and used in the analysis in later sections) are all dereddened. For the DR1 catalog, the correction is applied using the $E(B-V)$ values from the reddening map of \citet{Schlegel1998} and extinction coefficient $R_b$ derived using the \citet{Fitzpatrick1999} reddening law and the~\citet{Schlafly2011} adjusted reddening normalization parameter; for Y2Q catalog, the correction is made using the stellar locus regression. Details on the reddening corrections can be found from the corresponding references, \citet{desdr1} for DR1 and \citet{dw15b} for Y2Q.

\section{RESULTS}
\label{sec:results}
In this section, we describe the new stream members identified from this work, from which we derived the velocity gradient of the stream as well as the metallicity dispersion. We also compared our measurements and results with those from~\citet{tuc3} for the members in the Tuc~III core. 

\subsection{Spectroscopic Membership Determination}
\label{sec:membership}

We identified a total of 31 members in the AAT and IMACS combined dataset, 9 of which are members in the Tuc~III core and were previously confirmed in~\citet{tuc3}. The other 22 members are in the Tuc~III tails and are identified for the first time in this work. Adding the 26 members from the Tuc~III core from ~\citet{tuc3}, the total sample in the Tuc~III stream increases to 48 stars. As shown in Figure~\ref{fig:targets}, the member stars form a clear peak around $\vhel\sim-100~\kms$ and $\feh\sim-2.5$. Though a few candidate stars are at a similar velocity or metallicity if we only consider one property or the other, combining the two quantities separates the Tuc~III members from non-members, as shown in the lower right panel of Figure~\ref{fig:targets}. 

We note that the membership identification of Tuc~III stream members is a subjective selection process using the following parameters: velocity, metallicity, color, magnitude, and the spectrum itself. Specifically, we examined the candidate stars that have velocities between $-140\kms$ and $-70\kms$ star-by-star and listed their parameters in Table~\ref{tab:tuc3_spec}. Since the bulk velocity of the Tuc~III stream ($\sim-100~\kms$) is far away from that of the Milky Way disk stars, the Milky Way foreground contamination in this velocity range is minimal. The membership of the Tuc~III stream is mostly unambiguous, especially through their location on the velocity versus metallicity plot as shown in  Figure~\ref{fig:targets}. 
{For stars with low S/N spectra and no available EW measurements, we check with the best-fit RV template on the velocity measurements to assess whether the stars are metal-poor or not.}
We assess the member and non-members in the Tuc~III stream subjectively in this section.
In~\S\ref{sec:mixture}, we discuss the membership probability of all the observed stars using an objective Bayesian approach.

Among the 31 member stars in the Tuc~III stream, 29 were observed by AAT and 20 of them are the first identified members in the tidal tails. Two of them are BHB members (DES\,J234654.06$-$594331.7, DES\,J235248.09$-$602054.9) and one is an RHB member (DES\,J000529.05$-$600323.4), and the remaining stars are on the RGB. The RHB star shows no velocity variation from multiple measurements over several nights of AAT observations in 2016, so we conclude it is not an RR Lyrae star. We note that RHB stars are not common for dwarf galaxies at this luminosity and age, but from the velocity it is consistent with the Tuc~III member stars. Including or excluding this star does not change the kinematic properties of the stream. 

In order to check for possible binary motions of the stream member stars, we combined the spectra from the 2016 and 2017 AAT runs separately to provide independent measurements with a time baseline of $\sim13$~months for each star. We note that only the member stars that are both observed in \code{AAT-Field-1} and \code{AAT-Field-3} have repeated measurements after 13 months (see Figure~\ref{fig:targets}).
{We find that RGB member DES\,J234350.83$-$593925.6 is likely a binary, because we measured a radial velocity of $\vhel = -122.2\pm0.8~\kms$ during the 2016 run (average MJD = 57589) and $\vhel = -99.4\pm1.4~\kms$ on Aug 22 2017 (MJD = 57987). Its measured velocity from the combined spectra ($\vhel = -115.5 \pm 0.8~\kms$) is very similar to other stream members nearby; furthermore, it also has a very low metallicity ($\feh = -2.8 \pm 0.2$). We therefore conclude that it is a stream member. However, we exclude this star from the kinematic analysis in later sections.} All remaining AAT members do not show large velocity variations with the data we have obtained.

In contrast to the AAT, IMACS probes deeper but with a much smaller field of view. One subgiant (DES\,J235425.88$-$594103.3) and one MSTO star (DES\,J235435.00$-$593946.0) were uniquely detected by IMACS in one of the masks. In addition, RGB member DES\,J235349.12$-$593245.4 was observed by both AAT and IMACS and shows no velocity difference between the two. 

{We then discuss the non-members that have velocities close to the systemic velocity of the stream. Specifically, we list all of the non-member stars which are in the velocity range of $-140\kms$ and $-70\kms$: }
\begin{itemize} 
    \item A few candidate stars, including\\ DES\,J234319.89$-$592540.8, DES\,J234437.03$-$595405.6, DES\,J235853.63$-$595952.2, DES\,J000344.78$-$600048.2, DES\,J000758.93$-$594729.2, have relatively large CaT EWs, suggesting that they are non-members.
    \item DES\,J234949.16$-$602020.5 has very low S/N so no EW is measured, but a metal-rich RV template was selected as the best fit template.\footnote{The membership of this star will be discussed further in~\S\ref{sec:mixture}.} 
    \item DES\,J235855.20$-$591242.4 has a small CaT EW but its velocity is more than 20\kms off from the bulk velocity of the stream at its location. The independent RV measurements from AAT and IMACS are consistent within the 1$\sigma$ uncertainty, suggesting that it is not in a binary system. 
    \item DES\,J235158.27$-$591210.9 is similar. It also has a small EW but a large velocity offset ($> 20~\kms$) from the member stars at a similar location on the sky.
    \item DES\,J000104.89$-$594814.4 has a small EW and its velocity is about 10~\kms off from the bulk velocity of the stream at the location. Furthermore, its position on the color-magnitude diagram (CMD) is slightly offset from the other members. 
\end{itemize}
\noindent
{We suggest that for the last two non-members mentioned above, i.e. DES\,J235158.27$-$591210.9 and DES\,J000104.89$-$594814.4, more observations would be useful to check the possibility that they could be stream members with velocities offset from the stream by binary orbital motions.}

\begin{figure}[th!]
\plotone{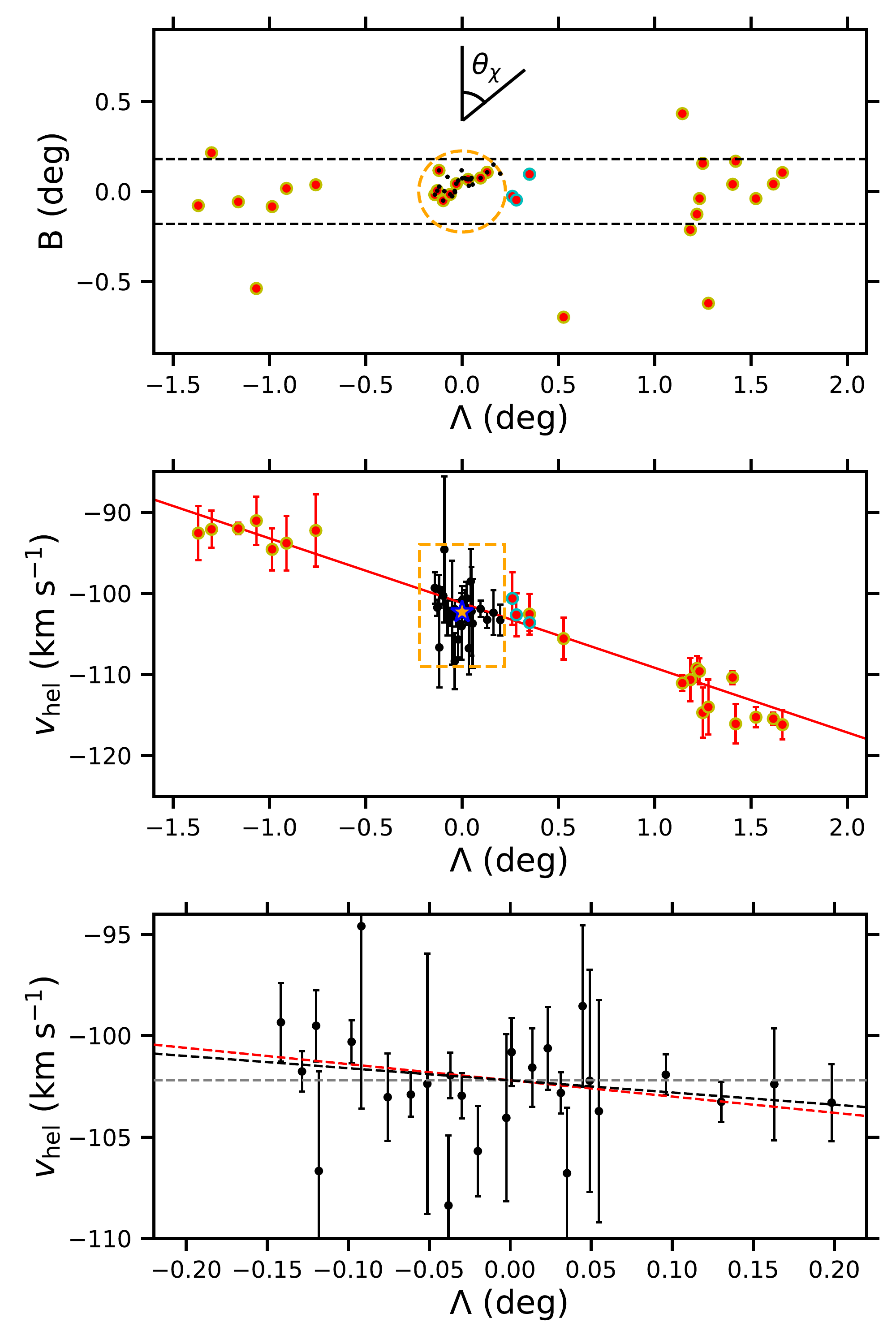}
\caption{\textbf{Top:} Members confirmed in this work in Tuc~III stream coordinates ($\Lambda$, B), where the center of Tuc~III core is at ($\Lambda, \Beta$) = (0, 0). Symbols are the same as in Figure~\ref{fig:targets}. Also plotted in black dots are the member stars of the Tuc~III core from~\citet{tuc3}, 9 of which overlap with the AAT confirmed members. The black dashed line shows the width $\sigma_w = 0.18\degr$ (1$\sigma$) of the stream from~\citet{shipp2018}. The orange dashed circle shows the definition of members in Tuc~III core ($r < 2.2r_h$) where all the members confirmed in~\citet{tuc3} are encircled. The definition of position angle $\theta_\chi$ is also illustrated. 
\textbf{Middle:} Heliocentric velocity $\vhel$ as a function of stream longitude $\Lambda$ for 22 members in the tidal tails. RGB member DES\,J235349.12$-$593245.4 was observed by both AAT (yellow circle) and IMACS (cyan circle); both measurements are presented. The red line shows the best fit velocity gradient and systemic velocity from the MCMC fit using the velocities of the tail members only. {The black circles represents the velocities of 26 core members from ~\citet{tuc3}.} The orange star symbol shows the systemic velocity of the Tuc~III core measured by~\citet{tuc3}. The uncertainty is smaller than the size of the symbol. 
\textbf{Bottom:} {A zoom in of the dashed orange rectangle in the middle panel,} in which the velocities of the 26 members from~\citet{tuc3} are also presented. The black dashed line shows the best fit velocity gradient from these 26 core members while the red dashed line indicates the gradient from the tails (i.e., the same as the red solid line in the middle panel) and the gray dashed line indicates a no gradient model. The gradient derived from core members alone ($-6.0\pm3.9~\kms\,\mathrm{deg}^{-1}$) is similar to what is detected in the stream ($\vgrad~\kms\,\mathrm{deg}^{-1}$), but with a much larger uncertainty. Due to the relatively large velocity uncertainties and small velocity differences observed in the core, the gradient in the core is statistically insignificant (see more discussion in the text).
}
\label{fig:vgrad}
\end{figure}

In order to derive the kinematic properties of the stream in the following sections, we transformed the member stars from celestial equatorial coordinates ($\alpha, \delta$) to the stream coordinates ($\Lambda, \Beta$) using Euler angles ($\phi,\theta,\psi = 264.23\degr, 120.29\degr, 267.51\degr$) where $\phi, \,\theta$ are derived from the pole of the Tuc~III stream from~\citet{shipp2018} by assuming a great circle orbit on the sky, and $\psi$ is chosen so that the center of the Tuc~III core has $\Lambda = 0$. We also list the transformation matrix in Appendix~\ref{sec:coords}. The member stars in stream coordinates ($\Lambda, \Beta$) are shown in the top panel of Figure~\ref{fig:vgrad}.

We observed candidate stars that are roughly $\pm$0.7\degr~from the Tuc~III stream track, taking into account the stream width of $\sigma_w = 0.13\degr$ (or FWHM = 0.3\degr) from~\citet{dw15b} at the discovery of the stream. The reanalysis from \citet{shipp2018} with improved datasets indicate the stream is slightly wider at $\sigma_w = 0.18\degr$. Surprisingly, we found three apparent member stars at least 3$\sigma$ away from $\Beta = 0$ (see Figure~\ref{fig:vgrad}), including one BHB member ($\Beta=-0.69\degr$), one RGB member ($\Beta =-0.692\degr$) and one RHB member ($\Beta =-0.54\degr$), respectively. {Finding 3 out of 22 members at $\Beta > 3\sigma_w$ may indicate that the stream profile is non-Gaussian. This can naturally arise if Tuc III is a globular cluster since while $\sim 5\%$ of stars will escape per relaxation time due to tidal stripping \citep{henon_1961}, $\sim1\%$ will escape from the core \citep[e.g.][]{spitzer_1987, baumgardt_etal_2002, alexander_gieles_2012} with a much larger velocity dispersion due to multi-body interactions. The $\sim 17\% $ stars ejected from the core would thus produce a much broader stream than the stars which were tidally stripped. Of course, this argument would not apply if Tuc III is a dwarf galaxy. }

{A non-Gaussian profile across the stream can also arise due to epicyclic motion along the stream \citep{kuepper_etal_2008,kuepper_etal_2010,Kuepper:2012}. In particular, away from pericenter the epicycles bunch up and can almost overlap \citep[see e.g. Fig. 9 of][]{Kuepper:2012}. This can create streams with stars significantly off the main track.}


{Alternatively, this may imply that one or some of these three stars are not true members of Tuc III stream, especially for the RHB member. As we note earlier, since the RHB star does not seem to be an RR Lyrae star, it is very uncommon for dwarf galaxies at this luminosity to have an RHB member. Proper motion from \gaia Data Release 2 (DR2)\footnote{See \url{https://www.cosmos.esa.int/web/gaia/dr2} for more details} can further assess the membership of these stars. 
}

\citet{shipp2018} estimate that the stellar mass of the stream is about $\stellarmass~\msun$. 
Assuming a \citet{2001ApJ...554.1274C} initial mass function with an age of 12.5 Gyr and metallicity of $\feh = -2.3$, we estimate a total of $\sim30\pm5$ member stars brighter than $g\sim19.5$ in the stream from 100 realizations of dwarf galaxy stellar populations randomly sampled using \code{ugali}\footnote{\url{https://github.com/DarkEnergySurvey/ugali}}. As a comparison, among all the confirmed members from the tails and the core, 26 members have $g<19.5$. 
{For most of the AAT spectra, we get S/N$\sim7$ at $g\sim19.5$; except for stars that were uniquely observed in the field of \code{AAT-Field3} where only a 2 hr exposure were taken, we get S/N$\sim4$ at $g\sim19.5$. About 90\% of the target stars that are brighter than $g\sim19.5$ have successful RV measurements. Those 10\% unsuccessful measurements are mostly from spectra in \code{AAT-Field3}. 
We conclude that our sample is near complete at $g<19.5$ for \code{AAT-Field1} and \code{AAT-Field2}, which covers roughly 3.6$\degr$ in total. }
However, the actual stream length is about $4.8\degr$ 
and therefore, we expect several additional brighter members near the ends of the stream that were not observed in this work.

\citet{dw15b} found that the stellar mass of the Tuc~III dwarf galaxy candidate is around $800\msun$, 
which is about 21\% of the total stellar mass of the stream. Assuming a Plummer profile, the enclosed stellar mass within two half-light radii ($2r_{h}$), close to the definition of the Tuc~III core here, is about 80\% of the total mass of the dwarf galaxy, i.e. 17\% of the total stellar mass of the stream.  As a comparison, for all the members at $g<19.5$, we identified 6 members in the core and 20 members in the tails, which confirms that the member stars in the core account for $6/26\sim23\%$ of the total members in the stream.  {As we expect additional brighter members would be found near the ends of the stream as discussed above, this ratio would become lower after finding more tail members.}

\subsection{Velocity Gradient}\label{vdisp}

\begin{figure*}[th!]
\plotone{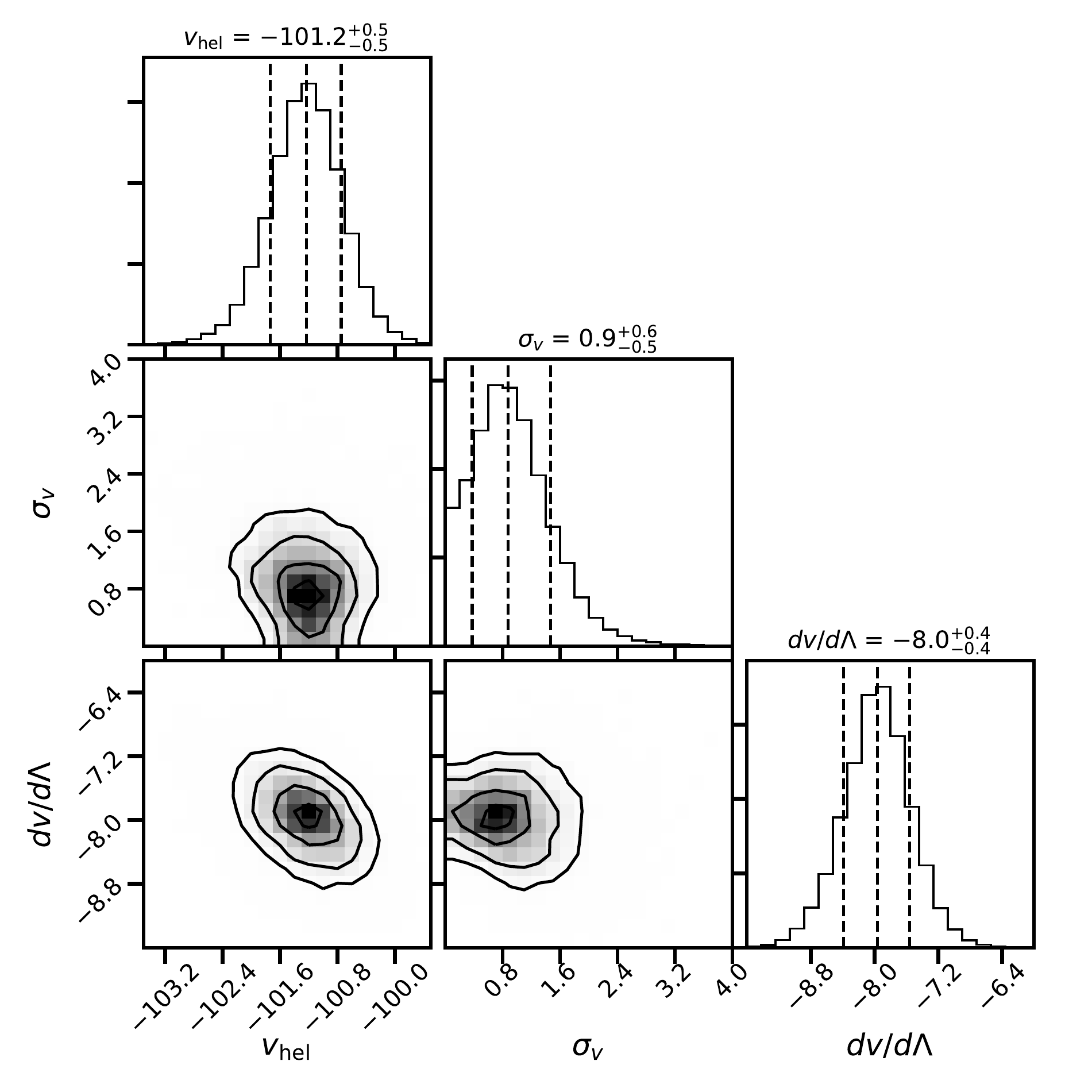}
\caption{Two-dimensional and marginalized posterior probability distribution from an MCMC sampler using a 3-parameter likelihood model. The three parameters are the systemic velocity $\vhel$ (in \kms) at the center of Tuc~III core (i.e. $\Lambda = 0$), velocity dispersion $\sigma_v$ (in \kms), and the velocity gradient along stream longitude $dv/d\Lambda$ (in $\kms\,\mathrm{deg}^{-1}$), respectively. Dashed lines in the 1-D histograms indicate the 16th, 50th, and 84th percentiles of the posterior probability distributions. A large velocity gradient of  $\vgrad~\kms\,\mathrm{deg}^{-1}$ is detected.}
\label{fig:mcmc}
\end{figure*}

As shown in Figure~\ref{fig:targets}, a clear velocity gradient, with increasing velocity towards larger right ascension ($\alpha_{2000}$, or equatorial east), is present in the Tuc~III stream. {In this section, we calculate the systemic velocity and velocity gradient using the tail stars measured in this paper, and compare the systemic velocity with what was derived in \citet{tuc3}. We also perform the fit with different setup to check if the results changes with different fitting parameters and different datasets. The results are summarized in Table~\ref{tab:vdisp_compare}.}

We first fit the RVs ($v$) and RV uncertainties ($\delta_v$) of 21 tail members (after excluding one probable binary member) via a maximum likelihood approach and a 3-parameter likelihood function similar to~\citet{eri2} to derive the systemic velocity \vhel~of the stream, i.e. the system velocity at the center of the Tuc~III core ($\Lambda = 0$), the velocity gradient along stream longitude $dv/d\Lambda$ , and the velocity dispersion $\sigma_v$:

\begin{equation}\label{mle_vdisp}
\medmuskip=-1.5mu
\thinmuskip=-1mu
\thickmuskip=-1mu
\log \mathcal{L} = -\frac{1}{2} \left [  \sum_{i = 1}^{N}\log (\sigma_{v}^2 + \delta_{v_i}^2) +  \sum_{n = 1}^{N} \frac{(v_i - v_{\rm hel} - \frac{dv}{d\Lambda}\Lambda_i)^2}{\delta_{v_i}^2 + \sigma_{v}^2} \right ].
\end{equation}\label{eq:mcmc_gradient}

Since \citet{tuc3} used flat priors to fit the systemic velocity and velocity dispersion, we use flat priors for all three parameters to have a direct comparison with~\citet{tuc3} later.\footnote{The overall posterior distribution will be smaller for the velocity dispersion if the Jeffreys prior is used instead~\citep[see, e.g.,][]{kim15_peg3, Li2018}} The posterior distribution from the MCMC sampler is shown in Figure~\ref{fig:mcmc}, and the best fit values are:

\begin{align*}
\vhel_{|(\Lambda=0)} & = \vbulk~\kms \\
d\vhel/d\Lambda          & = \vgrad~\kms\,\mathrm{deg}^{-1}\\
\sigmav              & = \vdisp~\kms
\end{align*}

\noindent
where we report the 50th percentile of the posterior and the uncertainty is calculated from the 16th and 84th percentiles. 
At a distance of 25~\kpc, the velocity gradient of $d\vhel/d\Lambda  = \vgrad~\kms\,\mathrm{deg}^{-1}$ corresponds to $18.3~\kms\,\kpc^{-1}$ projected on the sky {in heliocentric frame}. 
The best fit velocity gradient and systemic velocity, along with the velocities of the tail members are shown in the {middle} panel of Figure~\ref{fig:vgrad}. 

\begin{figure*}[th!]
\plotone{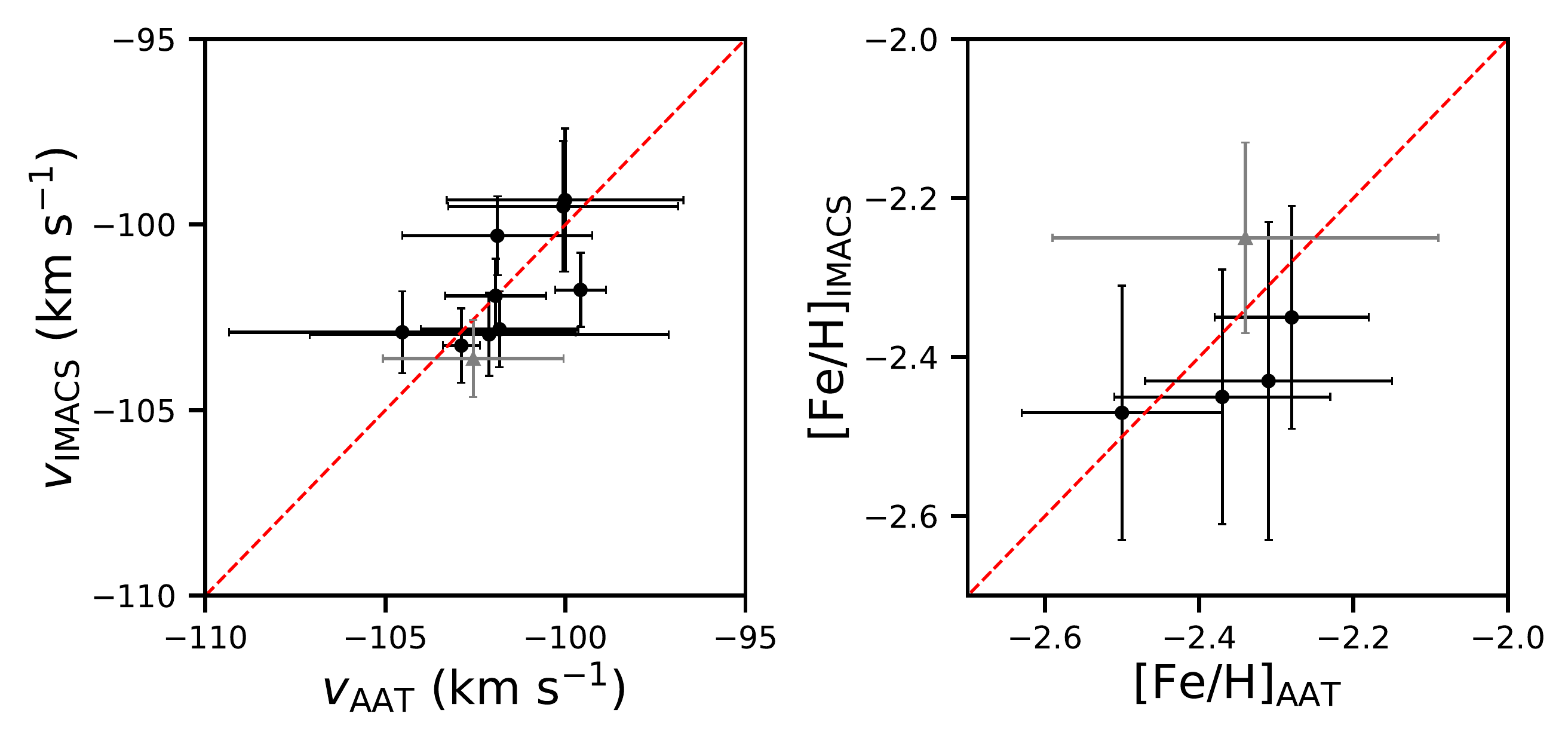}
\caption{A comparison of the measurements of RVs (left) and metallicities (right) from AAT and IMACS. Except for star DES\,J235349.12$-$593245.4 (coded as a gray triangle), the other IMACS measurements were taken from~\citet{tuc3}. The measurements from the two different instruments are consistent within 1$\sigma$ uncertainty for the majority of stars except for the RV of DES\,J235738.50$-$593611.7. The RV uncertainties from AAT measurements on most stars are larger mainly due to the lower S/N of the spectra.}
\label{fig:compare_stars}
\end{figure*}

We note that this is a fit using only the 21 out of 22 members in the tails (one excluded {because of apparent binarity}). We did not use the other 9 member stars in the core that were both measured in this work as well as in~\citet{tuc3}. Therefore, the fit gives an independent check on the systemic velocity of the Tuc~III core. The systemic velocity of the Tuc~III core is $-102.3 \pm 0.4~\kms$ ~\citep{tuc3}, from a sample of 26 core members. The difference between the two, $\sim1\kms$, is about 1.5$\sigma$ of the joint uncertainty. 
To test the origin of this slight velocity difference, we compare the individual member stars observed with both AAT and IMACS~\citep[mostly from][]{tuc3}, as shown in Figure~\ref{fig:compare_stars}. All stars have consistent velocity measurements within 1$\sigma$ uncertainty except for DES\,J235738.50$-$593611.7, {for} which the difference is {$\sim2\sigma$}. If we use only the 9 core members that are measured in both works, we get $\vhel = -101.6 \pm 0.5~\kms$ from AAT measurements and  $\vhel = -102.0 \pm 0.4~\kms$ from ~\citet{tuc3}. We conclude that the systematic offset between the two instruments is minimal. This is consistent with our comparison of the measurements from these two instruments in~\citet{Li2018}. This also confirms that these 9 core members do not show any binary motions. 

We also measured the velocity gradient and dispersion with 21 tail members plus 9 core members measured in this work. The results are consistent with fitting the tail sample alone.

We note that we fit the gradient along $\Lambda$ (i.e. $\Beta = 0$) in the earlier analysis. Similar to~\citet{eri2}, we introduce an additional degree of freedom on position angle $\theta_\chi$ {which is defined to be North-to-East in the stream coordinates} (see illustration in the top panel of Figure~\ref{fig:vgrad}) and run a 4-parameter fit in stream coordinates to check the possibility that the velocity gradient $dv/d\chi$ is not aligned with the stream. We found very similar results (see Table~\ref{tab:vdisp_compare}) as those from the 3-parameter fit, and $\theta_\chi= \PA$ is consistent with the case where the gradient is aligned with the stream ($\theta_\chi=  90\degr$).

\subsubsection{Equilibrium in the Tuc III core?}

{Enlightened by the fact that Tuc~III had a very close pericenter passage (see details in discussions in~\S\ref{sec:orbit} as well as in~\citealt{Erkal2018}), we examine the state of equilibrium in the Tuc III core and, specifically, we search for signatures of velocity gradient in the core.}
We applied the same 4-parameter fit to the 26 core members measured in~\citet{tuc3}.\footnote{For stars that have multiple measurements, the one with highest S/N was used.} We found a velocity gradient of $dv/d\chi = -6.7 \pm 6.1~\kms\,\mathrm{deg}^{-1}$ and a position angle of $\theta_\chi = 103\degr^{+42\degr}_{-58\degr}$ for the Tuc~III core, which is consistent with the gradient in the Tuc~III stream. However, the large uncertainty on both the gradient and the position angle indicate that the velocity gradient is poorly detected.
If we apply a 3-parameter fit instead, the corresponding gradient is similar with slightly smaller uncertainty, at $dv/d\Lambda = -6.0 \pm 3.9~\kms\,\mathrm{deg}^{-1}$. In the bottom panel of Figure~\ref{fig:vgrad}, we show the velocities of the 26 core members in stream coordinates along with the best fit gradient from the 3-parameter fit.
To assess the significance of the velocity gradient model in the Tuc~III core we compute the (logarithmic) Bayes' factor ($\ln{\rm B}$) comparing the gradient model with the null model (no gradient). (See \citealt{2008ConPh..49...71T} for a review of the Bayes' Factor and Bayesian model selection.)  We find that $\ln{\rm B} = -1.6$ for both the 3-parameter and 4-parameter models.  Values $>0$ ($<0$) favor (disfavor) the gradient model and values within the following ranges $0<1<2.5<5$ ($0>-1>-2.5>-5$) correspond to insignificant, low, moderate and significant evidence in favor (disfavor) of the gradient model \citep[based on the Jeffreys scale, see Table 1 of ][]{2008ConPh..49...71T}.  With only the core data, the gradient model is disfavored compared to the {no-gradient} model {at low statistical significance}. 
{We note that,} even if the core exhibits the same gradient as the stream, a $\sim0.3\degr$ extension in $\Lambda$ in the core will only have $\sim 2~\kms$ difference between the two ends, which is similar  to the uncertainty in the velocity measurements for individual stars. The Tuc~III velocity gradient therefore only becomes statistically significant once a large radial extent is observed. If more precise velocity measurements are obtained for stars in the range of $0.05\degr < \lvert  \Lambda \rvert\ < 0.5\degr$ (or more members are found in this range), it may be possible to identify the location of the transition between the remaining progenitor and the tidal tails. {The transition radius can be further compared with the tidal radius at the pericenter of Tuc III's orbit.}

\begin{figure*}[th!]
\plotone{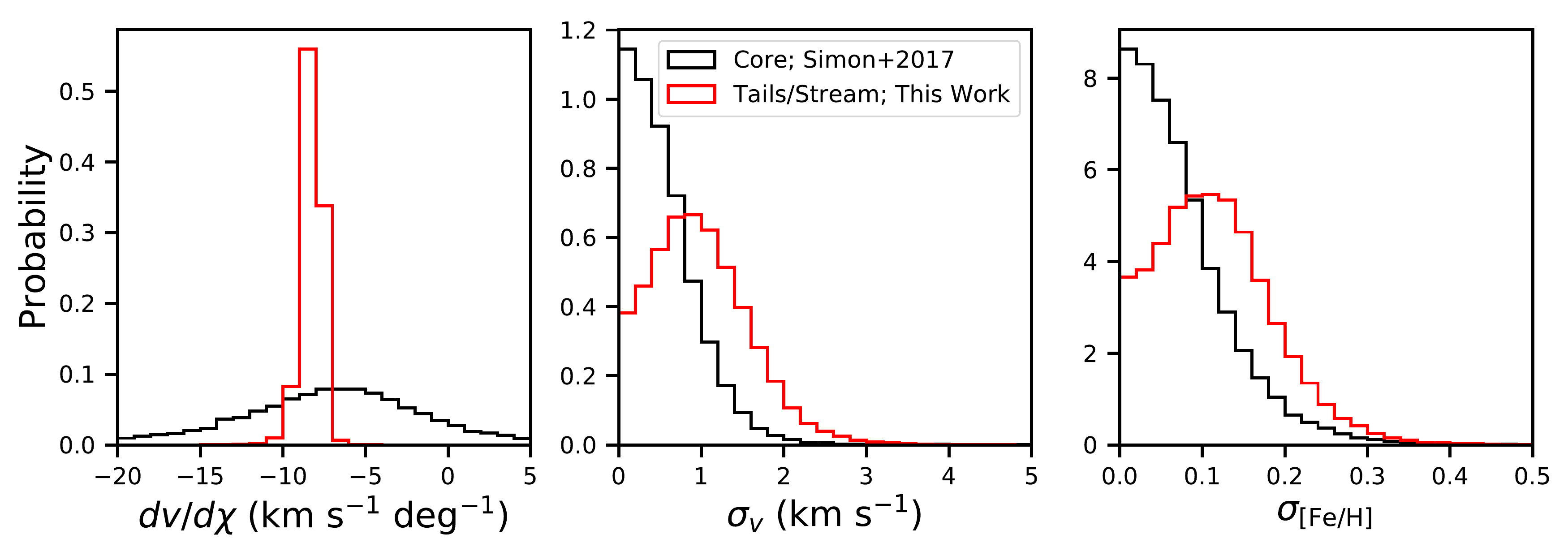}
\caption{A comparison of the velocity gradient $dv/d\chi$ (left), velocity dispersion $\sigma_v$ (middle), and metallicity dispersion $\sigma_\feh$ (right) derived from this work (red) on the Tuc~III streams and from~\citet{tuc3} (black) on the Tuc~III core. For velocity gradient and velocity dispersion, 21 tail members (one excluded due to binarity) from this work were used. For metallicity dispersion, 16 tail+core RGB members from this work were used.}
\label{fig:compare}
\end{figure*}

A comparison of the posterior distributions of the velocity gradient and velocity dispersion of the Tuc~III tails using 22 tail members from this work and those of the Tuc~III core using 26 core members are shown in the left and middle panel of Figure~\ref{fig:compare}. For both datasets, the 4-parameter fit described above is used.
In \citet{tuc3}, the velocity dispersion of the Tuc~III core was not resolved (i.e., $\sigma < 1.5~\kms$ at 95\% confidence level). As shown in the middle panel of Figure~\ref{fig:compare}, the velocity dispersion of the tails is likely slightly higher than the core, though the posterior distributions of the two largely overlap. The larger velocity dispersion in the tails may be a natural consequence of the ongoing tidal disruption.

\subsection{Metallicity and Metallicity Dispersion}
\label{sec:feh}

Among the 31 confirmed member stars in the Tuc~III stream, we obtained the metallicity of 16 RGB members, 4 of which are the brightest RGB members in the Tuc~III core and are also measured by~\citet{tuc3} with IMACS. One tail member was measured both by AAT and IMACS in this work. A comparison of the AAT measurements and IMACS measurements for these 5 members is presented in Figure~\ref{fig:compare_stars} and shows that there is no systematic offset between the two. The brightest core member (DES\,J235532) was also observed by~\citet{Hansen2017} with high resolution spectroscopy and the measured metallicity ($\feh = -2.25\pm0.18$) is comparable to what is measured in this work ($\feh = -2.28\pm0.10$).  We note that although more RGB members in the Tuc~III core have metallicity measurements in~\cite{tuc3}, we decide to only use the members measured from this work for the analysis of metallicity properties so that the limiting magnitude for both core members and tail members is relatively uniform ($g\lesssim$ 19).

\begin{figure}[th!]
\plotone{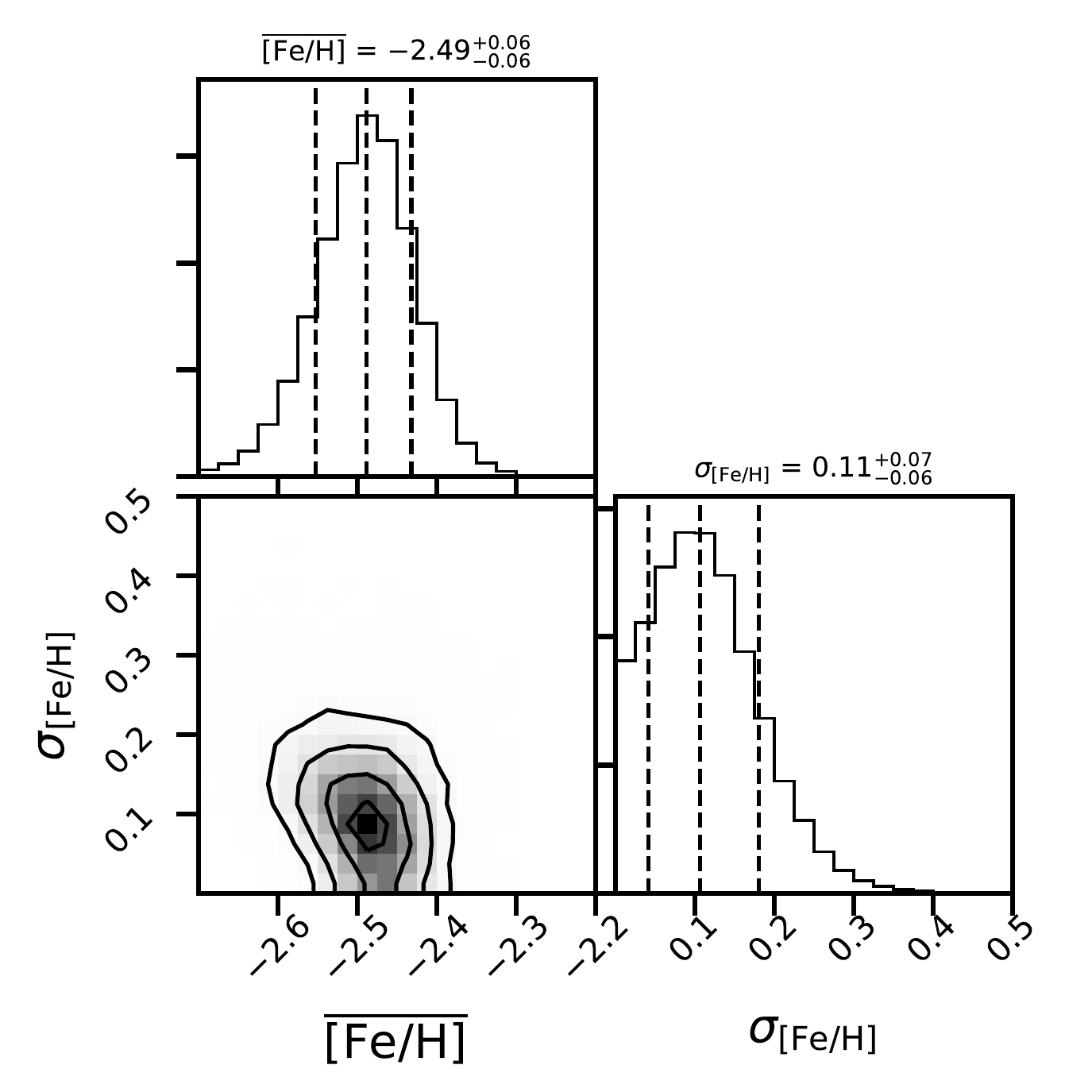}
\caption{Posterior distribution of mean metallicity $\overline{\feh}$ and metallicity distribution $\sigma_\feh$ from 16 RGB members.}

\label{fig:feh}
\end{figure}

\begin{figure}[th!]
\plotone{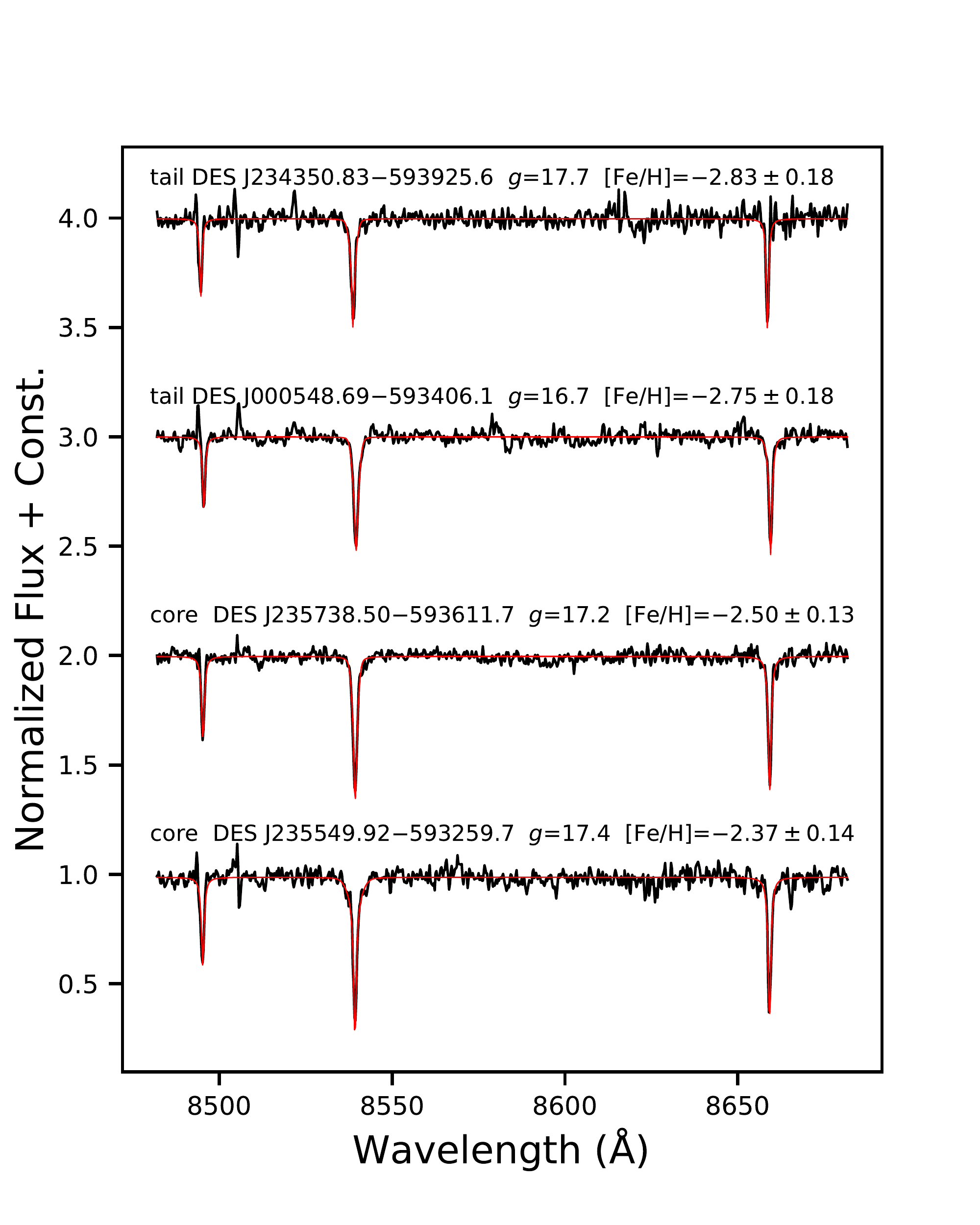}
\caption{The spectra of four Tuc~III stream members, two tail members and two core members, observed by AAT, shown in black lines. The red lines are the best fit model for measuring the CaT EWs, as described in~\S\ref{sec:measure}. The member stars are chosen so that they have similar brightnesses to minimize the surface gravity effect on CaT EWs. The two tail stars has smaller EWs and therefore are more metal-poor compared to the core stars at the similar brightness.}
\label{fig:spec}
\end{figure}

The metallicity of the 16 RGB members from this work spans from $\feh = -2.3$ to $\feh = -3.0$, as shown in Table~\ref{tab:tuc3_spec}. We found a mean metallicity of $\overline{\feh} = \fehmean$ and a metallicity dispersion of $\sigmafeh = \fehdisp$, with the posterior distribution presented in Figure~\ref{fig:feh}. As a comparison, \citet{tuc3} measured a mean metallicity of $\overline{\feh} = -2.44^{+0.07}_{-0.08}$ and an upper limit on the metallicity dispersion of $<0.19$ at 95.5\% confidence level for the Tuc~III core.
Similar to the velocity dispersion, the metallicity dispersion from the stream (core+tail) is slightly larger than that in the core (see right panel of Figure~\ref{fig:compare}). The increase in the dispersion might be a hint that the progenitor of the Tuc~III stream is more likely to be a dwarf galaxy rather than a star cluster (see discussions in \S\ref{sec:nature}). This dispersion is mainly driven by the three most metal-poor RGB members in the stream ($\feh < -2.7$). A comparison of their spectra to the core members are displayed in Figure~\ref{fig:spec}. Interestingly, these three most metal-poor RGB members are also among the farthest stream members from the Tuc~III center along the stream, as shown in the left lower panel in Figure~\ref{fig:targets}. 

\subsection{Spectroscopic Membership Probability}
\label{sec:mixture}

\begin{figure*}[th!]
\plotone{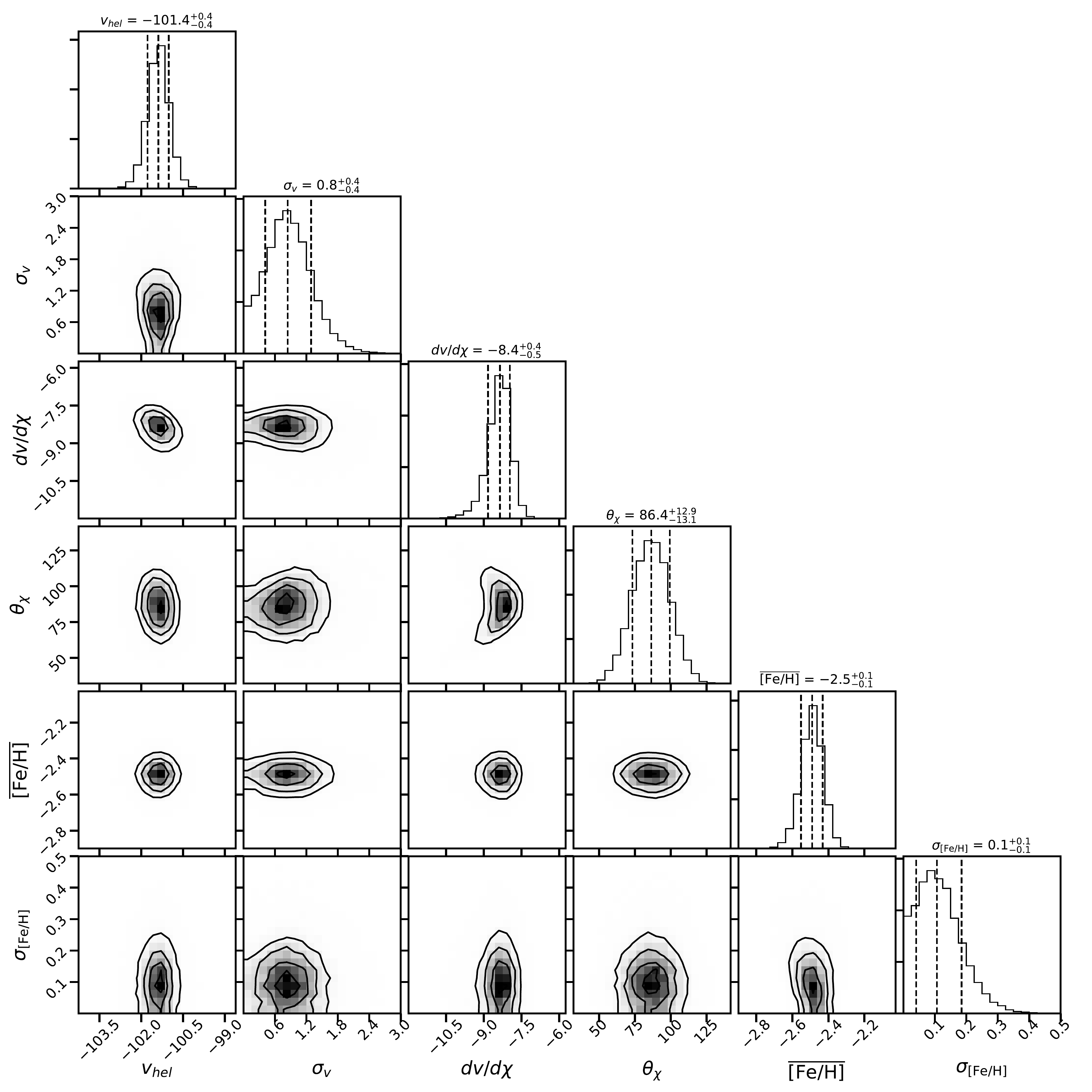}
\caption{Posterior distributions of the Tuc~III stream and Milky Way mixture model from 552 stars in the three AAT fields.  Only the Tuc~III stream properties of the mixture model are displayed here.  They are from left to right: systemic velocity (\vhel), velocity dispersion ($\sigma_v$), velocity gradient ($dv/d\chi$), position angle of the velocity gradient ($\theta_{\chi}$), average metallicity ($\overline{\feh}$) and, metallicity dispersion ($\sigma_\feh$).  The posteriors of  Tuc~III properties in the mixture model are very similar to the results from the subjective membership selection.}
\label{fig:mixture}
\end{figure*}

We construct a probabilistic mixture model as a cross-check of our membership selection determined in \S\ref{sec:membership} and to see if the exclusion of a Milky Way (MW) component adversely affected our results. We only apply the mixture model to the AAT data to consider a relatively uniform depth across the entire field.

The mixture model likelihood with the Tuc~III stream and MW components is written as:

\begin{equation}
\mathcal{P}_{\rm Total} = (1 - f_{\rm MW}) \mathcal{P}_{\rm Tuc \, III} + f_{\rm MW} \mathcal{P}_{\rm MW} \, . \\ 
\end{equation}

\noindent Where $f_{\rm MW}$ is the  fraction of stars in the MW population. We only use velocity and metallicity data in this mixture model with data vector $\mathcal{D}_i = (v_i, \delta_{v, i}, \feh_i, \delta_{\feh, i})$ where $v_i$ and $\feh_i$ are the velocity and metallicity of every observed star and $\delta_{v, i}$ and $\delta_{\feh, i}$ are the uncertainties of the measurements.  
Both velocity and metallicity are constructed with Gaussian distributions.  
We write the Gaussian distribution as:
 \begin{align*}
 \mathcal{N}(v,\sigma^2)=\frac{1}{\sqrt{2 \pi \sigma^2}}\exp{\left[-\frac{1}{2 }\frac{v^2}{\sigma^2}  \right] }
 \end{align*}
The MW component is:

\begin{align*}
\mathcal{P}_{\rm MW }(\mathcal{D}) &= \mathcal{N}(v_{\rm hel}^{\rm MW} - v_i,  (\sigma_{v}^{\rm MW} )^2 + \delta_{v, i}^2 ) \\
&\times \mathcal{N}(\overline{\feh}_{\rm MW} - \feh_i,  ( \sigma_\feh^{\rm MW} )^2 + \delta_{\feh, i}^2  ) \, . \\ 
\end{align*}

\noindent The Tuc~III stream model is similar but includes an additional term for the velocity gradient ($dv/d\chi$):

\begin{align*}
\mathcal{P}_{\rm Tuc \, III } &= \mathcal{N}(\vhel - v_i - \frac{d v}{d \chi} \chi_i  ,\sigma_{v}^2 + \delta_{v, i}^2 ) \\
 &\times \mathcal{N}(\overline{\feh} - \feh_i,\sigma_{\feh}^2 + \delta_{\feh, i}^2 ) \, . \\ 
\end{align*}

Overall, we have 11 free parameters:  Tuc~III stream ($\vhel$, $\sigmav$, $dv/d\chi$, $\theta_{\chi}$, $\overline{\feh}$, $\sigmafeh$), MW ($v_{\rm hel}^{\rm MW}$, $\sigma_{v}^{\rm MW}$, $\overline{\feh}^{\rm MW}$, $\sigma^{\rm MW}_{\feh}$), and $f_{\rm MW}$.  
For the low S/N stars without metallicity measurements, we average over all possible metallicity values; as the metallicity likelihood is normalized to one, this effectively excludes the metallicity term.
We assume linear priors for all parameters.
We determine the posterior distribution with the MultiNest package \citep{2008MNRAS.384..449F, 2009MNRAS.398.1601F}.
We compute membership probabilities ($p_i$) by computing the ratio of Tuc~III  likelihood to the total likelihood ($p_i = (1 - f_{\rm MW})\mathcal{P}_{\rm Tuc \, III}/\mathcal{P}_{\rm Total} $ ) and we refer to these as the Bayesian membership probabilities \citep{martinez11}.  The membership is computed for each point in the chain and the median value is adopted as the final $p_i$. 

We find 31 stars with non-zero membership ($p_i > 0.001$) and overall membership $\sum p_i = 28.3$. 
The 29 members (AAT only) in Section~\ref{sec:membership} all have $p_i >0.75$ and there are 2 stars previously considered non-members that have a non-zero membership in the mixture model.
The first non-member, DES\,J000104.89$-$594814.4, has $p_i=0.49$ and was considered a non-member previously due to the velocity offset and the offset in color from the Tuc~III CMD.  
Because the velocity offset from Tuc~III's velocity is small ($\sim10\kms$) and it is has a low metallicity, it has a non-zero membership in the mixture model.
The second, DES\,J234949.16$-$602020.5, has $p_i=0.73$ was considered a non-member due to the metal-rich template providing a better fit than the metal-poor template due to the large CaT EW.  
As the S/N was too low for an accurate CaT EW measurement, only the velocity was considered in the mixture model and the mixture model considers this star a probable member.
The properties of Tuc~III are not changed with respect to our results with the subjective analysis described in~\S\ref{sec:membership}; we conclude that our determination of the Tuc~III properties is robust.
Overall, our Bayesian membership probabilities agree with the subjective membership.  

We explored adding spatial information to the mixture model.
We precomputed spatial probabilities based on a simple Gaussian model in stream latitude $\Beta$ with stream width $\sigma_w = 0.18\degr$ \citep{shipp2018}.
{We find that the spatial probability lowers the membership probabilities of the candidate stars at larger $\Beta$ (especially the three members at $\Beta \gtrsim 3\sigma_w$), and therefore the overall membership decreases to $\sum p_i = 26.4$. However, adding spatial information does not change the posterior distribution of the kinematic and chemical properties of Tuc III stream as shown in Figure~\ref{fig:mixture}.}

\section{DISCUSSION}\label{sec:discuss}

\subsection{The properties of the Tucana III Stream}

\subsubsection{Density Variation along the Stream}\label{sec:density}

As shown in Figure \ref{fig:vgrad}, if we ignore the members confirmed by IMACS (which probes much deeper than AAT) and only focus on the 29 confirmed members from AAT, we notice obvious underdensities around $\Lambda\sim\pm0.5\degr$. As discussed in~\S\ref{sec:membership}, the bright members with $g < 19.5$ are mostly identified within the fields of 3 AAT pointings. We therefore believe this non-uniform distribution of bright member stars is not a cause of observational bias.  
Though these underdensities could be a result of small number statistics ($\sim 20$ tail stars from AAT), they may also just appear underdense relative to the epicyclic overdensities arising from tidal disruption \citep[e.g.][]{kuepper_etal_2008,kuepper_etal_2010} which have been seen in the Pal 5 stream \citep[e.g.][]{Kupper2017ApJ...834..112K,Erkal:2017}. 
Given the short amount of time needed to form a stream as long as Tuc~III~\citep[see][]{Erkal2018}, the fact that no more wiggles (overdensity + underdensity) have been seen in these bright members might also reflect the fact the progenitor has only had one pericentric passage (where the first tidal disruption happened). Additional modeling is needed to investigate the formation of the stream. Given the small sample with bright member stars, we suggest that the density variation along the stream longitude should be further investigated and verified with deeper photometry data.

\subsubsection{The Stream Orbit}\label{sec:orbit}

As shown in Figure~\ref{fig:targets} and \ref{fig:vgrad}, the radial velocity of the Tuc~III stream member stars decreases towards smaller right ascension ($\alpha_{2000}$) or larger stream longitude ($\Lambda$). {Tuc~III has a velocity of $v_\mathrm{GSR} = -195.2~\kms$ at $\Lambda = 0$ and $d v_\mathrm{GSR}/d\Lambda = -6.1~\kms\,\mathrm{deg}^{-1}$ in the Galactic Standard of Rest (GSR) frame. Therefore, the stream is moving towards the Galactic center; the west tail (or $\Lambda > 0$) is the leading arm and moving faster towards us, and the east tail (or $\Lambda < 0$) is the trailing arm and moving slower}. 

Furthermore, \citet{shipp2018} reported that a distance gradient of $\frac{d(m-M)}{d\Lambda} = 0.14 \pm 0.05 \magn \deg^{-1}$ was detected along the Tuc~III stream, implying (given the stream's position relative to the Galactic center) that the Tuc~III stream is on a radial orbit. The large velocity gradient measured in this work matches with the picture of the large distance gradient from photometry measurements. 
In fact, Tuc~III is likely on a highly eccentric ($e \sim 0.9$), inclined orbit with a pericenter of several kpc from the Galactic Center, though this orbit largely depends on the mass of the Large Magellanic Cloud (LMC). 
We refer readers to~\citet{Erkal2018} for a more detailed modeling work on the orbit of the Tuc~III stream, which uses the stream track and distance measured from the DES photometry, as well as the velocity and velocity gradient from this work.

The orbit of the Tuc~III stream will be further constrained by the proper motions of the stream, which will soon be measured by the upcoming \gaia DR2. We computed the expected precision of the proper motions with which the Tuc~III stream will be measured using the spectroscopically confirmed members from this work (see details in Appendix~\ref{sec:pmprecision}). The projected precision of 0.04 mas/yr (or 5 km/s at 25 kpc) will place very tight constraints on the orbit of Tuc~III stream, and place further constraints on the mass of LMC~\citep[see][]{Erkal2018}.

\subsubsection{The Nature of the Progenitor}\label{sec:nature}

The nature of Tuc III is still under debate. Due to its low metallicity and size, \citet{tuc3} tentatively suggested that Tuc~III is the tidally-stripped remnant of a dark matter-dominated dwarf galaxy. Indeed, if the total stellar mass of the progenitor is the same as the Tuc~III stream as measured by ~\citet{shipp2018} (i.e. $\stellarmass~\msun$), it would lie directly on the metallicity-luminosity relation of dwarf galaxies \citep{kirby13a}. 
However, recent work by~\citet{Simpson2018} found a similarly low metallicity for the faint globular cluster ESO280-SC06, further blurring the boundary between dwarf galaxies and star clusters.

Apart from the metallicity, the large size of Tuc~III ($r_h\sim44$~pc) relative to the globular cluster population is another piece of evidence favoring a dwarf galaxy origin. If the progenitor is a star cluster, the unusually large size would presumably be a consequence of tidal stripping. However, even though stripping plus observational biases can potentially inflate the size of faint star clusters~\citep{Contenta2017}, the radius of Tuc~III is still large enough to be difficult to explain.
While it is known that compact globular clusters ($r \lesssim 5$~pc) can survive very close encounters with the center of the Milky Way \citep[e.g.,][]{Sohn2018}, comparable measurements are not available for low-mass dwarf galaxies or extended outer halo clusters.  Theoretical modeling of objects on such orbits could provide additional clues to the nature of Tuc~III.
For future studies, it is also important to obtain better and/or additional velocity measurements in the inner region of the stream (i.e. $0.05\degr < \lvert  \Lambda \rvert\ < 0.5\degr$) to detect where the gradient starts and to identify the location of the transition between the remaining progenitor and the tidal tails (see discussions in \S\ref{vdisp}).

As discussed in~\S\ref{sec:feh}, driven by the three most metal-poor star near the two ends of the stream, we found a marginally larger metallicity dispersion for the Tuc~III stream compared to the upper limit in the Tuc~III core \citep{tuc3}, suggesting a possible dwarf galaxy origin for the Tuc~III stream.
If the outer halo of the progenitor object was tidally stripped first, then seeing more metal-poor stars farther from the core (along the stream direction) indicates a possible metallicity gradient in the progenitor, where the metal-poor stars are less centrally concentrated than the metal-rich ones. Similar trends have been seen in other Milky Way satellite galaxies, but at larger stellar masses~\citep[see, e.g.,][]{2011ApJ...727...78K}. This could also explain why the metallicity dispersion of the Tuc~III core is small despite its progenitor being a dwarf galaxy. 
We measure $\sigma_{\feh}=0.11_{-0.06}^{+0.07}$ dex; as \citet{Willman2012} concluded that $\sigma_{\feh} > 0.2$ dex robustly diagnoses a dwarf galaxy, the metallicity dispersion does not definitely classify this object.  
If Tuc~III is a dwarf galaxy, the small dispersion inferred here is likely the result of the following two causes. 
First, as a result of the low S/N of the AAT spectra, the uncertainty on the CaT-derived \feh\ of individual stars is relatively large (most members have \feh\ uncertainties larger than the median metallicity dispersion of 0.11 dex). Therefore, despite a large metallicity range (between $-3.0 < \feh < -2.3$), the metallicity dispersion is not completely resolved (i.e., it is still consistent with zero). The dispersion can be refined with better metallicity determinations, either with high-resolution spectroscopic follow-up observations, or perhaps higher S/N CaT spectra. 
Second, the sample size of the most metal-poor population is small, i.e., only 3 members at $\feh < -2.7$. It is possible that the progenitor contained more metal-poor members at $\feh < -2.7$ but they were stripped first and are now outside the known extent of the stream. Testing the metallicity gradient hypothesis requires mapping the entire Tuc III stream, in particular seeking lower-metallicity stars that might be located near the ends of the stream.

Furthermore, the chemical abundance patterns of the member stars could help the classification. For example, light-element abundance correlations (e.g Na-O, Na-Al, Mg-Al) appear to be ubiquitous in star clusters \citep[e.g.,][]{Johnson2017ApJ...842...24J, Bastian2017arXiv171201286B}. Many confirmed Tuc~III stream members from this work are bright enough for a detailed abundance analysis via high-resolution spectroscopic observations.

Furthermore, based on the width of the stream, \citet{shipp2018} derived the progenitor mass to be $\sim8\times10^4$~\msun. If the stellar luminosity of $2.8\times10^3$~\lsun (or $M_v = -3.8$) reported in ~\citet{shipp2018} is close to the total stellar luminosity of the progenitor, it would imply a mass-to-light ratio of $\sim40~\msun/\lsun$, indicating a possible dwarf galaxy classification, though most of the ultra-faint dwarf galaxies at a similar luminosity have a much larger mass-to-light ratio\footnote{We also note most of the mass-to-light ratios are defined within the half-light radius, which is very different from how the progenitor mass was calculated based on the width of stream.}. 

\subsection{Comparison with other streams and satellites}

\subsubsection{Palomar 5}
Of all the thin streams known so far, the tidal tails of the globular cluster Palomar 5 (Pal 5) are in many ways similar to the Tuc~III tails -- both have an unambiguous progenitor identified and a similar velocity dispersion (see below). 
First detected in SDSS \citep{Odenkirchen2001}, the stellar stream of the Palomar 5 globular cluster extends over at least 22$^{\circ}$ \citep{Ibata2016ApJ...819....1I}. A velocity gradient {(in heliocentric frame)} between $0.4-1.0~{\rm km \, s^{-1} \, deg^{-1}}$ was first detected by \citet{Odenkirchen2009AJ....137.3378O} and later confirmed with larger data sets \citep{Kuzma2015MNRAS.446.3297K, Ibata2017ApJ...842..120I}. While at a similar heliocentric and Galactocentric distance, it is noteworthy that Tuc~III possesses a velocity gradient 10$\times$ larger than that of Pal 5.
\citet{Ibata2017ApJ...842..120I} find that the stellar mass of the tidal tails is $3\times$ the mass of the core, which is very similar to the Tuc~III stream, though Tuc~III appears shorter on the sky (5$^{\circ}$ vs. 22$^{\circ}$), partially due to the projection from its orientation.

Though the progenitor of the Tuc~III stream is more likely to be a dwarf galaxy based on the large range of metallicities in member stars as discussed in \S\ref{sec:nature}, the velocity dispersion of the Tuc~III tails is smaller than that of the Pal 5 stream~\citep[$2.1\pm0.4~\kms$;][]{Kuzma2015MNRAS.446.3297K}. This low velocity dispersion of the Tuc~III stream makes it a good target to search for stream density perturbations caused by close encounters with dark matter subhalos \citep[e.g.][]{Erkal:2015b}. The short length of the Tuc~III stream, however, may imply that the stream formed recently and therefore there might not have been enough time for dark matter subhalos to perturb the stream density.  A more precise model of the stream will clarify the extent to which the stream is only apparently short due to it being aligned with our line of sight from the Sun. Even if the short length is only a projection effect, it will likely make it harder to search for gaps and wiggles along the stream. In addition, if there are density variations near the progenitor due to its secular disruption, these will need to be accounted for in the search for subhalos.

\subsubsection{Tidal features associated with dwarf galaxies}

The kinematics and morphology of the Tuc~III tails leave no doubt that they 
are physically associated with the satellite and that the tidal stream contains a large velocity gradient.  This result suggests that the observation of velocity gradients can be a good way to assess the dynamical state of dwarf galaxies \citep{Piatek95}.  Below we discuss other Milky Way dwarfs that have been claimed to contain extra-tidal features and/or velocity gradients and compare them with the Tuc~III stream.

Circumstantial photometric and kinematic evidence has been used to argue that several other dwarf galaxies are being tidally disrupted.  For example, unusually high ellipticities \citep[e.g., Hercules, Ursa Major II;][]{Sand2009ApJ...704..898S, Munoz2010AJ....140..138M}, irregular outer isophotes \citep[e.g., Ursa Major I, Ursa Major II;][although see \citealt{Martin2008} regarding the significance of such features]{Okamoto2008A&A...487..103O, Munoz2010AJ....140..138M}, extra-tidal sub-structures \citep[e.g., Hercules;][]{Sand2009ApJ...704..898S}, and kinematic sub-structure or velocity gradients \citep[e.g., Coma Berenices, Hercules, Leo V, Ursa Major II;][]{sg07, Aden2009ApJ...706L.150A, Collins2017MNRAS.467..573C} have been found in several satellites and interpreted as tidal features.  It is important to keep in mind, however, that the common attribution of such features to tidal stripping is not borne out by simulations of the stripping process \citep{Munoz2008}. 

A prime example of a dwarf galaxy often suggested to be disrupting is Hercules.
Many authors have considered its extremely elongated stellar distribution as evidence of tidal disruption \citep{Belokurov:2007, Coleman2007ApJ...668L..43C, Sand2009ApJ...704..898S, Roderick2015ApJ...804..134R}. 
Extra-tidal stellar overdensities, especially along the major axis, have also been identified \citep{Sand2009ApJ...704..898S, Fabrizio2014A&A...570A..61F, Roderick2015ApJ...804..134R}, and several RR Lyrae variables are located at large projected separations from the dwarf \citep{Garling2018ApJ...852...44G}. 
Note that while multiple studies have detected stellar overdensities, many of them do not overlap.
Hercules has also been claimed to contain a velocity gradient \citep{Aden2009ApJ...706L.150A, Deason2012MNRAS.425L.101D}, but the statistical significance of the gradient is very low (1.2$\sigma$) and a much larger spectroscopic sample over a wider area would be needed to test its reality (similar to our results in the Tuc III core). \citet{Martin2010ApJ...721.1333M} argued that Hercules could be an unbound stellar stream resulting from the disruption of a dwarf galaxy.
\citet{Kupper2017ApJ...834..112K} suggested that Hercules is on a very eccentric orbit and that Hercules and any extra structure is perpendicular to the orbit. 

Of the ultra-faint dwarf galaxies often cited as undergoing tidal disruption in the literature, Leo V is notable for its similarities to Tuc~III. 
There is some evidence for tidal disruption based on the stellar distributions \citep{Belokurov2008ApJ...686L..83B, deJong2010ApJ...710.1664D, Sand2012ApJ...756...79S}, in particular the extended BHB population \citep{Belokurov2008ApJ...686L..83B, Sand2012ApJ...756...79S} and RR Lyrae stars \citep{Medina2017ApJ...845L..10M}.
Leo V has a tentative velocity gradient  \citep{Collins2017MNRAS.467..573C}, with roughly four times the magnitude of Tuc~III ($\sim 80 \kms \,{\rm kpc}^{-1}$) but it was measured with only 8 stars over just $\sim 3\arcmin$. 
Similarly, \citet{Walker2009ApJ...694L.144W} find two potential members at large radii ($r\approx 13\arcmin$) and argue that Leo V is losing mass.  
A velocity gradient and members at large radii could indicate stripping from Leo V.
If Leo V is undergoing tidal disruption similar to Tuc~III, the lack of apparent tails may be due to the large distance of the satellite. In this scenario, deeper observations of the main sequence might reveal an extended structure. As a note, the surface brightness of Tuc~III core is about 29~\magn~arcsec$^{-2}$~\citep{dw15b} while the surface brightness of the tails is about 3~\magn~arcsec$^{-2}$ fainter~\citep{shipp2018}.

A key difference between Tuc~III and Hercules and Leo~V is that the latter two are quite distant satellites of the Milky Way ($d > 130$~kpc; \citealt{Musella2012, Sand2012ApJ...756...79S}), well beyond the region where the Milky Way's tidal field could be causing stripping.  If these objects are on very highly eccentric orbits then they could have suffered significant tidal stripping at pericenter and are now located near apocenter \citep{Kupper2017ApJ...834..112K}, but the required orbital eccentricities to bring them within a few kpc of the Galactic center at pericenter are extreme ($e > 0.95$).
Moreover, numerical simulations indicate that dwarfs that are not completely disrupted should quickly return to equilibrium after a pericentric passage \citep[e.g.,][]{Penarrubia2008,Penarrubia2009,Kazantzidis2011,Barber2015}, in which case tidal features are not expected to be seen for objects that are currently far out in the halo of the Milky Way.  While it has been suggested that Leo~V could be physically associated with its neighbor Leo~IV and that a tidal interaction between the two is possible \citep{deJong2010ApJ...710.1664D}, the mass required for the two systems to be gravitationally bound to one another is implausibly large for their luminosities \citep{deJong2010ApJ...710.1664D,Blana2012}.

Regarding the ultra-faint dwarf Segue 2, \citet{kirby13a} argued that it was tidally stripped because it does not lie on the stellar mass-metallicity relationship.  A similar argument was made for the Tuc~III core~\citep{tuc3}. Including the stellar mass in the tidal tails will move Tuc~III into agreement with other dwarf galaxies with respect to the stellar mass-metallicity relationship.  In contrast with Tuc~III, there are not clear tidal features seen in the Segue 2 stellar distribution, even with data that reaches the main-sequence of Segue 2 \citep{Belokurov2009MNRAS.397.1748B}. 

\section{Selecting Metal-poor RGB Stars with DES Photometry}
\label{sec:colorcolor}

\begin{figure*}[th!]
\plotone{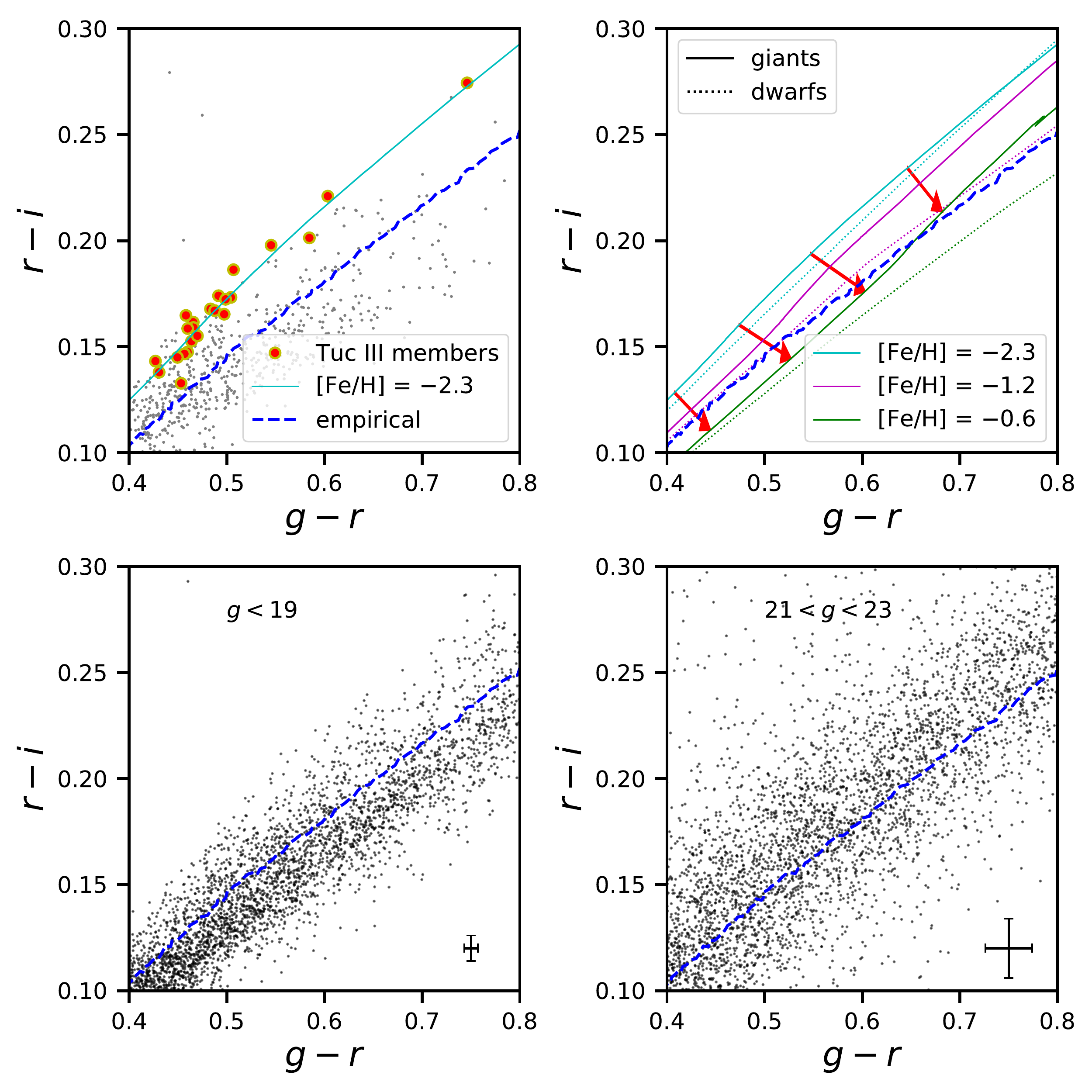}
\caption{\textbf{Top left:} Color-color diagram in $g-r$ vs. $r-i$ for the observed targets (gray dots) and confirmed members (red filled circles) in the Tuc~III stream. The dashed blue line is an empirical stellar locus of DES photometry. All confirmed Tuc~III members lie on one side of the empirical stellar locus and align well with a metal-poor isochrone at $\feh = -2.3$ using the synthetic magnitude from Dotter isochrones. \textbf{Top right:} Synthetic magnitude from Dotter isochrones at age = 12.5 Gyr and at various metallicities, along with the empirical stellar locus as shown in the top left panel. The synthetic magnitude from the isochrone indicates a strong metallicity dependent on $g-r$ color at a given $r-i$ color. Also plotted red vectors are the isotherm lines (or blanketing vector) for giant stars using the synthetic magnitude from the Dotter isochrones. At a given stellar temperature, the $g-r$ color increase (i.e. redder) and $r-i$ color decrease (i.e. bluer) from a metal-poor stellar population to a metal-rich stellar population.
\textbf{Bottom:} Stars in the field of the Tuc~III stream with $g<19$ (bottom left) and $21 < g < 23$ (bottom right). The brighter (fainter) stellar bin is dominated by nearby disk stars (distant halo stars) and has more metal-rich (metal-poor) stars, thus the majority stars are below (above) the stellar locus.  The high precision DES photometry could provide a rough metallicity estimation of red stars ($0.4 < g-r < 0.8$) based on the $g-r$ vs. $r-i$ color of the stars, and could further provide an estimation of metallicity distribution function of the Milky Way. We note that in the bottom right panel, due to the poorer star-galaxy separation at the fainter magnitude, we expect some galaxy contaminations whose colors are far away from the stellar locus. The median photometric uncertainty for each magnitude bin is also shown. The uncertainties are computed as a quadrature sum of the statistical uncertainty from DES DR1 catalog  (\code{WAVG\_MAGERR\_PSF}) and the systematic uncertainty as reported in DES DR1~\citep{desdr1}. In the brighter bin, the uncertainty is dominated by the systematic precision at 6--7 mmag (note the spectroscopically confirmed Tuc~III stream members have similar uncertainties). In the fainter bin, the uncertainty is dominated by the \code{WAVG\_MAGERR\_PSF} quantity with a median at 0.01-0.02 mag.
}
\label{fig:color-color}
\end{figure*}

Although halo substructures such as stellar streams and dwarf galaxies are most commonly identified by their MSTO stars, which dominate the total stellar counts of a system for typical survey depths, the spectroscopic follow-up observations for membership identification and kinematic measurements are mostly performed on RGB stars due to the faintness of the MSTO stars. In contrast to the hundreds to thousands of MSTO members, there are usually only a few dozens of RGB members in a stellar stream or an ultra-faint dwarf galaxy. The efficiency of membership identification is extremely low, due to a large amount of contamination from foreground stars in the Milky Way disk and the low density of RGB members in these substructures, especially for stellar streams where the surface density is much lower compared to dwarfs. For example, of the 552 stars for which we obtained successful velocity measurements with the AAT observations, only 29 are members of the Tuc~III stream.  
Fortunately, stellar streams and dwarf galaxies are mostly old and metal-poor populations, and therefore, if the stellar metallicity can be roughly estimated using the photometry, the foreground contaminants could be largely removed, increasing the success rate of follow-up spectroscopy.

The broadband colors of stars are sensitive to their chemical composition. For example, many studies have found correlations between stellar colors and metallicities for M-dwarfs~\citep{Lepine2008AJ....135.2177L,Bochanski2013AJ....145...40B, Lepine2013AJ....145..102L,Li2014AJ....148...60L}. 
For F/G stars, \citet{Ivezic2008ApJ...684..287I} presented a correlation to estimate their effective temperature and metallicity using the position of the stars on the SDSS $u-g$ vs $g-r$ diagram. This method was based upon the traditional ultraviolet (UV) excess method or line-blanketing effect~\citep[see, e.g.,][]{Wildey1962, Sandage1969}. In other words, the metallicity of a subdwarf (i.e., metal-poor dwarf star) can be estimated with the difference between the star's $U-B$ color and that which would be measured for a more metal-rich star with the same $B-V$ color, because more metal lines are present in the shorter wavelengths of a stellar spectrum.
Unfortunately, DES does not routinely use the DECam $u$-band. Here, we study the locations of confirmed member stars in the Tuc~III stream on the $g-r$ vs $r-i$ diagram. Thanks to the high photometric precision of DES, the method described below can be used to improve the target selections for future spectroscopic observations seeking members in streams and dwarf galaxies.

We constructed a color-color diagram for the confirmed RGB members in the Tuc~III stream in $g-r$ vs $r-i$ using the DES DR1 photometry when available.\footnote{Stars that have the Y2Q photometry in Table~\ref{tab:tuc3_spec} are not included in Figure~\ref{fig:color-color}.} We focus on the color range of $0.4 < g-r < 0.8$ because this is the range for the RGB member stars in Tuc~III stream where the foreground contamination dominates.  

As shown in the top panel of Figure~\ref{fig:color-color}, all confirmed Tuc~III members lie on one side of the empirical stellar locus, which was constructed as the median of stars in the dereddened DES photometry. 
{Specifically, we select a sample of stars over the full survey footprint in regions with low interstellar extinction -- $E(B-V) < 0.015$ using the reddening map of \citep{Schlegel1998} -- that are unsaturated and measured with high S/N in each of the $g$, $r$, $i$, $z$ bands (e.g., $16 \lesssim r \lesssim 21$). We bins these selected stars ($\sim 200,000$) according to their $g-z$ color with $\sim 500$ stars in each of 429 bins and evaluate the median stellar colors in each bin.}
As the stream members are all metal-poor stars ($\feh < -2$), the clumping of the members suggests that the color of these RGB stars slightly depends on the metallicity of the star.  We therefore plot the Dotter isochrones at age = 12.5 Gyr and various metallicities in the top right panel of Figure~\ref{fig:color-color} and find a clear indication that at a given $r-i$ color, metal-poor stars tend to be bluer in $g-r$. 
This trend is very similar to the $u-g$ vs $g-r$ diagram as seen in the Figure 2 of~\citet{Ivezic2008ApJ...684..287I}. Similar to the UV excess in $U-B$ (or $u-g$ for SDSS), metal-poor stars also present a $g-r$ excess at a given $r-i$ color. We also plot the blanketing vectors~\citep[see, e.g.,][]{Sandage1959,Wildey1962} in the $g-r$ vs $r-i$ diagram, which shows the shift in position from a metal-poor giant star to a metal-rich giant star at a constant temperature using the synthetic magnitude from the Dotter isochrones. For a more metal-rich star, the $g-r$ color gets redder and the $r-i$ color gets bluer.

This metallicity-dependent color will improve the efficiency of selecting stream or dwarf RGB candidate members by at least a factor of 50\% (because all the member stars of the Tuc~III stream are on one side of the stellar locus). This is extremely valuable for the spectroscopic follow-up program where the number of fibers or slitlets of a multi-object spectrograph in one exposure is limited, and especially useful for stellar stream follow-up where the member stars are sparsely populated and the foreground contamination from Milky Way disk stars is relatively high. Furthermore, this color-color selection can remove foreground metal-rich disk stars and improve the detection significance for distant substructure searches (i.e. dwarfs and streams) using photometry alone since in these distance structures ($d > 200~\kpc$), only RGBs are brighter than the limiting magnitude of the imaging survey. 

Note that from $\feh = -2.3$ to $\feh = -0.6$, the difference in $r-i$ is less than 0.05 mag. Therefore, this color difference cannot be revealed without the high-precision DES photometry~\citep[rms $<$ 0.01 mag; see][]{Burke:2017}. Furthermore, this color-color selection is most efficient for the brightest stars where the uncertainty from photon noise is negligible ($g<21$). Fortunately, our spectroscopic targets are usually bright RGB stars and therefore we can take advantage of the precise photometric calibration of DES.

Following the trend discussed above, photometric metallicity in principle could be derived statistically for stars at $0.4 < g-r < 0.8$ based on their colors, and thereby derive the metallicity distribution function of the Milky Way's disk and its stellar halo, though a more sophisticated calibration is needed to derive a more precise correlation between the metallicity of the stars and their positions in the $g-r$ vs. $r-i$ diagram. As a proof of concept, we plot stars in the region of the Tuc~III stream without any prior color-magnitude cut in the bottom panels of Figure~\ref{fig:color-color}. We selected stars in two groups, $g < 19$ and $ 21 < g < 23$. The brighter stars are nearby and therefore dominated by metal-rich disk stars and the fainter stars are more distant and therefore dominated by metal-poor halo stars. The brighter group has a majority of its stars below the empirical stellar locus (and therefore is more metal-rich) while the fainter group has more stars above the locus (and therefore is more metal-poor). We leave a more thorough study on this topic to a future paper. 

\section{SUMMARY}
\label{sec:summary}

We report on a spectroscopic analysis of the Tuc~III stream using the 2dF+AAOmega spectrograph on the Anglo-Australian Telescope and the IMACS spectrograph on the Magellan/Baade Telescope. We identify, for the first time, 22 members in the tidal tails of Tuc~III. Together with the 26 members in the Tuc~III core previously confirmed by~\citet{tuc3}, this study yields a total sample of 48 stars in the Tuc~III stream. 
Using the tail members, we measured a large velocity gradient of $\vgrad~ \kms$\,deg$^{-1}$ along the stream, consistent with the picture of the large distance gradient detected from the DES photometry \citep{shipp2018}.
This velocity gradient, many times larger than that of the Pal~5 stream, for instance, strongly suggests that Tuc~III is on a radial orbit and passed close to the Galactic center. 
The membership and velocity information obtained in this work allow a detailed, precise orbit of Tuc~III to be constructed~\citep[see][]{Erkal2018}, which will further our understanding of the mass distribution of our Galactic neighborhood, including the relative roles that dark matter, disk stars and the LMC play in determining overall halo dynamics.

We found several more metal-poor member stars near the ends of the stream. These more metal-poor members farther from the center of Tuc III result in a slightly larger metallicity dispersion for the stream than that for the core alone as derived in~\citet{tuc3}, indicating that the progenitor of the Tuc~III stream is likely to be a dwarf galaxy rather than a star cluster. However, the metallicity dispersion we found is still smaller than most dwarf galaxies at a similar luminosity. Additional metal-poor members farther from the center of the stream may be found in future observations if such a metallicity gradient is genuine. 

We found that in a color-color diagram of $g-r$ vs. $r-i$, all the member stars in the Tuc~III stream are systematically redder in $r-i$ color (or bluer in $g-r$) than most non-member stars. The high precision of DES photometry allows us to identify metal-poor stars photometrically. This metallicity-dependent color offers a more efficient method for selecting metal-poor targets and will increase the efficiency of selecting stream members for future spectroscopic follow-up programs. Furthermore, the color-color selection can eliminate foreground metal-rich disk stars and improve the detection significance in finding dwarf galaxies and stellar streams using DES data (or other imaging surveys with a similar or better photometric precision).

\acknowledgements{
TSL thanks Jo Bovy and Sergey Koposov for helpful conversations. 
JDS acknowledges support from the National Science Foundation under grant AST-1714873. 
ABP acknowledges generous support from the George P. and Cynthia Woods Institute for Fundamental Physics and Astronomy at Texas A\&M University. 
{DE thanks Mark Gieles for helpful discussions.} DE acknowledges financial support from the European Research Council under the European Union's Seventh Framework Programme (FP/2007- 2013) / ERC Grant Agreement no. 308024.  
EB acknowledges financial support from the European Research Council (StG-335936). {The authors thank the anonymous referee for careful reading of the manuscript and providing useful comments.}

Based on data obtained at Siding Spring Observatory via program A/2016A/26.
We acknowledge the traditional owners of the land on which the AAT stands, the Gamilaraay people, and pay our respects to elders past and present. 

This research has made use of NASA's Astrophysics Data System Bibliographic Services.

This paper has gone through internal review by the DES
collaboration.

Funding for the DES Projects has been provided by the U.S. Department of Energy, the U.S. National Science Foundation, the Ministry of Science and Education of Spain, 
the Science and Technology Facilities Council of the United Kingdom, the Higher Education Funding Council for England, the National Center for Supercomputing 
Applications at the University of Illinois at Urbana-Champaign, the Kavli Institute of Cosmological Physics at the University of Chicago, 
the Center for Cosmology and Astro-Particle Physics at the Ohio State University,
the Mitchell Institute for Fundamental Physics and Astronomy at Texas A\&M University, Financiadora de Estudos e Projetos, 
Funda{\c c}{\~a}o Carlos Chagas Filho de Amparo {\`a} Pesquisa do Estado do Rio de Janeiro, Conselho Nacional de Desenvolvimento Cient{\'i}fico e Tecnol{\'o}gico and 
the Minist{\'e}rio da Ci{\^e}ncia, Tecnologia e Inova{\c c}{\~a}o, the Deutsche Forschungsgemeinschaft and the Collaborating Institutions in the Dark Energy Survey. 

The Collaborating Institutions are Argonne National Laboratory, the University of California at Santa Cruz, the University of Cambridge, Centro de Investigaciones Energ{\'e}ticas, 
Medioambientales y Tecnol{\'o}gicas-Madrid, the University of Chicago, University College London, the DES-Brazil Consortium, the University of Edinburgh, 
the Eidgen{\"o}ssische Technische Hochschule (ETH) Z{\"u}rich, 
Fermi National Accelerator Laboratory, the University of Illinois at Urbana-Champaign, the Institut de Ci{\`e}ncies de l'Espai (IEEC/CSIC), 
the Institut de F{\'i}sica d'Altes Energies, Lawrence Berkeley National Laboratory, the Ludwig-Maximilians Universit{\"a}t M{\"u}nchen and the associated Excellence Cluster Universe, 
the University of Michigan, the National Optical Astronomy Observatory, the University of Nottingham, The Ohio State University, the University of Pennsylvania, the University of Portsmouth, 
SLAC National Accelerator Laboratory, Stanford University, the University of Sussex, Texas A\&M University, and the OzDES Membership Consortium.

Based in part on observations at Cerro Tololo Inter-American Observatory, National Optical Astronomy Observatory, which is operated by the Association of 
Universities for Research in Astronomy (AURA) under a cooperative agreement with the National Science Foundation.

The DES data management system is supported by the National Science Foundation under Grant Numbers AST-1138766 and AST-1536171.
The DES participants from Spanish institutions are partially supported by MINECO under grants AYA2015-71825, ESP2015-66861, FPA2015-68048, SEV-2016-0588, SEV-2016-0597, and MDM-2015-0509, 
some of which include ERDF funds from the European Union. IFAE is partially funded by the CERCA program of the Generalitat de Catalunya.
Research leading to these results has received funding from the European Research
Council under the European Union's Seventh Framework Program (FP7/2007-2013) including ERC grant agreements 240672, 291329, and 306478.
We  acknowledge support from the Australian Research Council Centre of Excellence for All-sky Astrophysics (CAASTRO), through project number CE110001020, and the Brazilian Instituto Nacional de Ci\^encia
e Tecnologia (INCT) e-Universe (CNPq grant 465376/2014-2).

This manuscript has been authored by Fermi Research Alliance, LLC under Contract No. DE-AC02-07CH11359 with the U.S. Department of Energy, Office of Science, Office of High Energy Physics. The United States Government retains and the publisher, by accepting the article for publication, acknowledges that the United States Government retains a non-exclusive, paid-up, irrevocable, world-wide license to publish or reproduce the published form of this manuscript, or allow others to do so, for United States Government purposes.

}

{\it Facilities:} 
 \facility{Anglo-Australian Telescope (AAOmega+2dF); Magellan/Baade (IMACS).}
 
{\it Software:} 
\code{astropy} \citep{Astropy2013}, 
\code{corner.py} \citep{corner}, 
\code{emcee} \citep{Foreman_Mackey:2013},  
\code{matplotlib} \citep{Hunter:2007}, 
\code{numpy} \citep{numpy:2011}, 
\code{PyGaia}\footnote{\url{https://github.com/agabrown/PyGaia}},
\code{scipy} \citep{scipy:2001}, 
\code{ugali} \citep{Bechtol:2015}\footnote{\url{https://github.com/DarkEnergySurvey/ugali}}.

\bibliographystyle{apj}
\bibliography{main}{}

\clearpage

\appendix

\section{A. Coordinate Transformation Matrix}\label{sec:coords} 

In \S~\ref{sec:membership} we described the transformation from celestial coordinates ($\alpha, \delta$) to the stream coordinates ($\Lambda, \Beta$) for Tuc III stream using Euler angles ($\phi,\theta,\psi = 264.23\degr, 120.29\degr, 267.51\degr$), so that the stream is roughly aligned along $\Beta = 0$ and the Tuc~III core is at $\Lambda=0$. The transformation from ($\alpha, \delta$) to ($\Lambda, \Beta$) is given by

\begin{eqnarray}
\begin{bmatrix}
\cos(\Lambda) \cos(\Beta)\\
\sin(\Lambda) \cos(\Beta)\\
\sin(\Beta)
\end{bmatrix}&=&\nonumber\\
\begin{bmatrix}
0.505715 & -0.007435 & -0.862668\\
-0.078639 & -0.996197 & -0.037514\\ 
0.859109 & -0.086811 & 0.504377
\end{bmatrix} 
&\times &
\begin{bmatrix}
\cos(\alpha) \cos(\delta)\\
\sin(\alpha) \cos(\delta)\\
\sin(\delta)
\end{bmatrix}\nonumber
\end{eqnarray}

\section{B. Expected Proper Motion Precision from GAIA DR2 }\label{sec:pmprecision} 

The 29 member stars confirmed by AAT are relatively bright ($g<20)$ and will soon have proper motion measurements from \gaia DR2. Here, we estimate the expected precision of proper motion on the Tuc~III stream from \gaia DR2~\citep{gaiadr1}. We computed the expected proper motion uncertainties from \gaia DR2 for every star using the \code{PyGaia} package\footnote{\url{https://github.com/agabrown/PyGaia}} given the \gaia $G$-band magnitude and $V-I$ color of each star. We first convert the DES photometry to \gaia $G$-band photometry using the transformation equation reported in the Appendix of~\citet{desdr1}. Since $V-I$ color is unavailable for these stars, we replaced it with DES $g-i$ color instead. We note there is a small offset between the two, but a shift of 0.1~\magn in $V-I$ will only cause a 0.5~$\mu$as~yr$^{-1}$ change in the proper motion for the stars at $G=20$. We therefore conclude the effect of replacing $V-I$ with $g-r$ is minimal. Since proper motion errors scale like $t^{-1.5}$ where $t$ is the duration of observations, we also scaled the error by a factor of $4.5\times$ of what \code{pygaia} computes, taking into account that DR2 only includes the \gaia data from the first 22 months of the 5 year entire mission length\footnote{\url{https://www.cosmos.esa.int/web/gaia/dr2}}.
The projected proper motion uncertainty is roughly $0.1 < 0.2 < 0.5 < 1.0$~mas~yr$^{-1}$ for individual stars with $r$-band magnitude of $16.1 < 17.3 < 18.6 < 19.6 $.

We then compute the weighted averaged uncertainty of the 29 member stars as the expected precision of the proper motion on the Tuc~III stream, which is on the order of $\sim0.04$~mas~yr$^{-1}$. We caution that this projected overall uncertainty is calculated by assuming the uncertainty on each individual star is largely dominated by the statistical uncertainty and therefore the overall uncertainty will be reduced by averaging all of the measurements together. If the precision from \gaia is systematics-limited at this brightness, then the final overall uncertainty on Tuc~III could be much larger.

We also note that this weighted average uncertainty is mainly determined by the 6 brightest RGB members, which each have projected proper motion uncertainties of $<0.2$~mas~yr$^{-1}$. Recently, 4 RR Lyrae stars have been found (Mart{\'\i}nez-V{\'a}zquez in prep.) along the Tuc~III stream that are not in this spectroscopically confirmed sample. RR Lyrae stars in Tuc~III are relatively bright. Assuming a magnitude of $G\sim17.5$ and $V-I\sim0.2$, each RR Lyrae star will have a proper motion uncertainty of $0.2-0.3$~mas~yr$^{-1}$. Including these RR Lyrae stars will further improve the precision of the stream proper motion.

\section{C. Potential Members of the Tucana~III Stream}\label{sec:moremember} 

During the velocity measurements, we found another 52 candidate members that suggest a tentative velocity in the range of $-140~\kms < v < -70~\kms$. However, the low S/N ($1 \lesssim v \lesssim 6$) of the spectra did not pass the visual inspection in the fit. We did not include them in the analysis but we list these stars and the best-fit velocity in Table~\ref{tab:tuc3_spec_pot}. We caution the use of these measurements in RV (and therefore the uncertainties are not provided), but we suggest that these stars can be followed up spectroscopically with larger telescopes to verify their velocities and  membership. Some of these stars are brighter than the limiting magnitude of \gaia DR2, and accordingly, their membership status could be tested in the near future through precise proper motion measurements.


\clearpage
\clearpage

\begin{turnpage}

\begin{deluxetable*}{c r r c c c c r r c c c c c }
\tabletypesize{\scriptsize}
\tablecaption{Velocity and metallicity measurements of the observed stars.
\label{tab:tuc3_spec}
}

\tablehead{ID\tablenotemark{a} & $\alpha_{2000}$ & $\delta_{2000}$  & $g$\tablenotemark{b} & $r$\tablenotemark{b} & Cat\tablenotemark{b} & Inst\tablenotemark{c} & S/N & \multicolumn{1}{c}{$v$} & ${\rm EW}$ & ${\rm [Fe/H]}$ & Mem\tablenotemark{d} & Prob\tablenotemark{e} & Comment\\ 
 & (deg) & (deg) & (mag) & (mag) &  &   &  & \multicolumn{1}{c}{(\kms)} & (\AA) &  &  &  &}

\startdata
 \multicolumn{14}{c}{ \ruleline{Tucana III Leading/West Tail} } \\
DES\,J234319.89$-$592540.8 & 355.83288 & $-$59.42799 & 18.194 & 17.719 & DR1 & AAT & 22.2 & $-72.93 \pm 0.98$ & $5.01 \pm 0.36$ & \noinfo &  0 & 0.00 &     \\
DES\,J234327.66$-$593542.1 & 355.86526 & $-$59.59502 & 18.931 & 18.465 & DR1 & AAT & 12.9 & $-116.19 \pm 1.79$ & $1.30 \pm 0.32$ & $-2.97 \pm 0.29$ &  1 & 1.00 &     \\
DES\,J234350.83$-$593925.6 & 355.96177 & $-$59.65712 & 17.678 & 17.151 & Y2Q & AAT & 33.8 & $-115.46 \pm 0.76$ & $1.83 \pm 0.30$ & $-2.83 \pm 0.18$ &  1 & 1.00 &   binary  \\
DES\,J234434.87$-$594407.7 & 356.14531 & $-$59.73549 & 18.339 & 17.835 & DR1 & AAT & 20.8 & $-115.27 \pm 1.23$ & $2.15 \pm 0.49$ & $-2.51 \pm 0.27$ &  1 & 0.99  &     \\
DES\,J234437.03$-$595405.6 & 356.15431 & $-$59.90157 & 17.005 & 16.391 & DR1 & AAT & 40.7 & $-75.88 \pm 0.77$ & $5.49 \pm 0.27$ & \noinfo &  0 & 0.00  &     \\
DES\,J234521.78$-$593131.3 & 356.34076 & $-$59.52536 & 19.340 & 18.883 & DR1 & AAT &  9.6 & $-116.09 \pm 2.43$ & $1.65 \pm 0.52$ & $-2.62 \pm 0.36$ &  1 & 0.99  &     \\
DES\,J234531.06$-$593908.6 & 356.37941 & $-$59.65240 & 17.672 & 17.127 & DR1 & AAT & 36.0 & $-110.38 \pm 0.84$ & $2.27 \pm 0.38$ & $-2.59 \pm 0.20$ &  1 & 0.98 &     \\
DES\,J234642.46$-$601828.1 & 356.67692 & $-$60.30780 & 19.046 & 18.576 & DR1 & AAT &  9.8 & $-114.02 \pm 3.39$ & $2.01 \pm 0.49$ & $-2.44 \pm 0.34$ &  1 & 0.98 &     \\
DES\,J234642.83$-$593152.8 & 356.67844 & $-$59.53133 & 19.487 & 19.038 & DR1 & AAT &  8.7 & $-114.70 \pm 3.08$ & $1.85 \pm 0.79$ & $-2.47 \pm 0.52$ &  1 & 0.97  &     \\
DES\,J234654.06$-$594331.7 & 356.72524 & $-$59.72548 & 17.240 & 17.395 & DR1 & AAT & 21.7 & $-109.61 \pm 1.61$ & \noinfo & \noinfo &  1 & 0.91  &  BHB  \\
DES\,J234702.14$-$594843.2 & 356.75890 & $-$59.81200 & 18.459 & 17.967 & DR1 & AAT & 17.7 & $-109.26 \pm 1.52$ & $1.99 \pm 0.59$ & $-2.58 \pm 0.34$ &  1 & 0.99 &      \\
DES\,J234719.86$-$595348.0 & 356.83277 & $-$59.89668 & 19.449 & 18.989 & DR1 & AAT &  8.9 & $-110.65 \pm 2.68$ & \noinfo & \noinfo &  1 & 0.92 &      \\
DES\,J234727.52$-$591504.9 & 356.86466 & $-$59.25137 & 18.296 & 17.789 & DR1 & AAT & 23.1 & $-111.07 \pm 0.99$ & $2.23 \pm 0.43$ & $-2.48 \pm 0.24$ &  1 & 1.00 &      \\
DES\,J234949.16$-$602020.5 & 357.45482 & $-$60.33902 & 18.137 & 17.596 & DR1 & AAT &  6.2 & $-113.59 \pm 3.12$ & \noinfo & \noinfo &  0 & 0.74 &     \\
DES\,J235158.27$-$591210.9 & 357.99278 & $-$59.20303 & 17.541 & 17.271 & Y2Q & AAT & 31.7 & $-88.46 \pm 0.84$ & $4.98 \pm 0.28$ & \noinfo &  0 & 0.00 &     \\
DES\,J235248.09$-$602054.9 & 358.20036 & $-$60.34857 & 17.265 & 17.404 & DR1 & AAT & 16.3 & $-105.58 \pm 2.57$ & \noinfo & \noinfo &  1 & 0.90 &   BHB  \\
DES\,J235349.12$-$593245.4 & 358.45467 & $-$59.54593 & 18.524 & 18.025 & DR1 & AAT & 17.1 & $-102.56 \pm 2.51$ & $2.41 \pm 0.46$ & $-2.34 \pm 0.25$ &  1 & 0.98 &     \\
  &   &   &   &   &   & IMACS & 28.8 & $-103.61 \pm 1.04$ & $2.60 \pm 0.21$ & $-2.25 \pm 0.12$ &    & \noinfo &    \\
DES\,J235425.88$-$594103.3 & 358.60784 & $-$59.68424 & 20.288 & 20.008 & DR1 & IMACS &  6.8 & $-102.65 \pm 2.64$ & \noinfo & \noinfo &  1 & \noinfo &    subgiant  \\
DES\,J235435.00$-$593946.0 & 358.64582 & $-$59.66279 & 20.695 & 20.493 & DR1 & IMACS &  4.1 & $-100.63 \pm 3.24$ & \noinfo & \noinfo &  1 & \noinfo &  MSTO   \\
DES\,J235439.51$-$594118.7 & 358.66463 & $-$59.68854 & 19.997 & 19.611 & DR1 & IMACS &  9.7 & $-71.48 \pm 1.75$ & $1.87 \pm 0.29$ & \noinfo &  0 & \noinfo  &     \\
 \multicolumn{14}{c}{ \ruleline{Tucana III Core\tablenotemark{f} } } \\
DES\,J235532.68$-$593115.0 & 358.88616 & $-$59.52083 & 16.090 & 15.344 & DR1 & AAT & 125.7 & $-102.89 \pm 0.51$ & $3.75 \pm 0.22$ & $-2.28 \pm 0.10$ &  1 & 0.99 &     \\
DES\,J235549.92$-$593259.7 & 358.95799 & $-$59.54990 & 17.400 & 16.815 & DR1 & AAT & 22.0 & $-101.94 \pm 1.40$ & $2.84 \pm 0.29$ & $-2.37 \pm 0.14$ &  1 & 1.00 &      \\
DES\,J235620.76$-$593310.2 & 359.08651 & $-$59.55282 & 18.857 & 18.374 & DR1 & AAT & 11.6 & $-101.83 \pm 2.19$ & $2.33 \pm 0.27$ & $-2.31 \pm 0.16$ &  1 & 0.99 &      \\
DES\,J235650.50$-$593421.0 & 359.21042 & $-$59.57250 & 19.950 & 19.496 & DR1 & AAT &  6.0 & $-102.12 \pm 4.99$ & \noinfo & \noinfo &  1 & 0.83 &     \\
DES\,J235707.46$-$593743.0 & 359.28110 & $-$59.62861 & 19.727 & 19.296 & DR1 & AAT &  5.7 & $-104.53 \pm 4.82$ & \noinfo & \noinfo &  1 & 0.81 &      \\
DES\,J235726.04$-$593938.2 & 359.35851 & $-$59.66061 & 19.265 & 18.800 & DR1 & AAT & 13.2 & $-101.89 \pm 2.64$ & \noinfo & \noinfo &  1 & 0.89 &      \\
DES\,J235730.25$-$592930.7 & 359.37602 & $-$59.49186 & 19.342 & 18.878 & DR1 & AAT &  8.1 & $-100.06 \pm 3.19$ & \noinfo & \noinfo &  1 & 0.88 &      \\
DES\,J235738.50$-$593611.7 & 359.41040 & $-$59.60325 & 17.173 & 16.569 & DR1 & AAT & 45.9 & $-99.58 \pm 0.71$ & $2.66 \pm 0.25$ & $-2.50 \pm 0.13$ &  1 & 1.00 &     \\
DES\,J235745.46$-$593726.5 & 359.43941 & $-$59.62401 & 19.806 & 19.379 & DR1 & AAT &  8.2 & $-100.01 \pm 3.29$ & \noinfo & \noinfo &  1 & 0.88 &      \\
 \multicolumn{14}{c}{ \ruleline{Tucana III Trailing/East Tail} } \\
DES\,J235853.63$-$595952.2 & 359.72345 & $-$59.99784 & 18.956 & 18.467 & DR1 & AAT & 10.3 & $-121.84 \pm 2.14$ & $3.87 \pm 0.63$ & \noinfo &  0 & 0.00 &      \\
DES\,J235855.20$-$591242.4 & 359.73001 & $-$59.21177 & 18.701 & 18.219 & DR1 & AAT &  9.8 & $-78.54 \pm 2.86$ & $2.16 \pm 0.29$ & \noinfo &  0 & 0.00 &      \\
 &  &  &  &  &  & IMACS &  6.6 & $-74.02 \pm 1.52$ & \noinfo & \noinfo &   & \noinfo &     \\
DES\,J000104.89$-$594814.4 &   0.27039 & $-$59.80400 & 18.833 & 18.394 & DR1 & AAT &  8.0 & $-85.13 \pm 4.37$ & $2.19 \pm 0.60$ & \noinfo &  0 & 0.49 &     \\
DES\,J000234.74$-$593056.8 &   0.64473 & $-$59.51578 & 19.341 & 18.881 & DR1 & AAT &  5.7 & $-92.26 \pm 4.46$ & \noinfo & \noinfo &  1 & 0.78      \\
DES\,J000344.78$-$600048.2 &   0.93656 & $-$60.01339 & 19.109 & 18.670 & DR1 & AAT &  7.8 & $-129.28 \pm 3.42$ & \noinfo & \noinfo &  0 & 0.00 &      \\
DES\,J000347.26$-$593114.6 &   0.94693 & $-$59.52072 & 19.104 & 18.646 & DR1 & AAT &  7.1 & $-93.82 \pm 3.38$ & \noinfo & \noinfo &  1 & 0.86 &   \\
DES\,J000427.21$-$593648.3 &   1.11336 & $-$59.61342 & 18.771 & 18.283 & DR1 & AAT & 10.7 & $-94.58 \pm 2.58$ & \noinfo & \noinfo &  1 & 0.87 &     \\
DES\,J000529.05$-$600323.4 &   1.37102 & $-$60.05651 & 17.543 & 17.318 & DR1 & AAT & 21.6 & $-91.05 \pm 2.97$ & \noinfo & \noinfo &  1 & 0.85 &    RHB  \\
DES\,J000548.69$-$593406.1 &   1.45288 & $-$59.56835 & 16.770 & 16.123 & Y2Q & AAT & 40.4 & $-92.01 \pm 0.70$ & $2.34 \pm 0.35$ & $-2.75 \pm 0.18$ &  1 & 1.00 &     \\
DES\,J000639.63$-$591658.5 &   1.66513 & $-$59.28291 & 18.813 & 18.316 & DR1 & AAT &  9.8 & $-92.11 \pm 2.30$ & $2.09 \pm 0.27$ & $-2.46 \pm 0.16$ &  1 & 0.99 &      \\
DES\,J000727.58$-$593357.0 &   1.86493 & $-$59.56584 & 19.321 & 18.865 & Y2Q & AAT &  7.3 & $-92.58 \pm 3.35$ & \noinfo & \noinfo &  1 & 0.82 &      \\
DES\,J000758.93$-$594729.2 &   1.99552 & $-$59.79145 & 18.530 & 18.038 & DR1 & AAT & 11.7 & $-99.87 \pm 2.08$ & $5.83 \pm 0.58$ & \noinfo &  0 & 0.00 &    \\
\enddata
\tablecomments{Only stars with $ -140 < \vhel < -70 ~\kms$ are presented here. The remaining measurements are available in the online version in machine readable format.\\ 
(a) The star IDs are computed based on the RA ($\alpha_{2000}$) and Dec ($\delta_{2000}$) from either DES DR1 or Y2Q catalog in the format of DES\,Jhhmmss.ss--ddmmss.s. Due to the small difference in the astrometry between the two catalogs, the star ID in DR1 catalog could be slightly different (in the last decimal point) from the Y2Q catalog; the latter one was used to produce the star IDs in the previous literatures for Tuc~III~\citep[e.g.,][]{tuc3,Hansen2017}.\\
(b) Quoted magnitudes represent the dereddened PSF magnitude in either DR1 or Y2Q catalog.\\
(c) Telescopes and instruments used for obtaining the spectra of the targets. AAT stands for AAT/2df+AAOmega; IMACS stands for Magellan/IMACS. \\
(d) The membership of Tuc~III stream stars through a subjective membership identification. (See details in \S\ref{sec:membership}.) Mem = 1 are members; Mem = 0 are non-members.\\
(e) The membership probability of Tuc~III stream stars through an objective mixture model. Measurements from the AAT observations are used for modeling. (See details in \S\ref{sec:mixture}.) \\
(f) These 9 Tuc~III core members are also observed by ~\citet{tuc3}.
}
\end{deluxetable*}

\end{turnpage}

\clearpage

\begin{deluxetable*}{lccccc}
\setlength{\tabcolsep}{0.2in}
\tabletypesize{\scriptsize}
\tablecolumns{6}
\tablewidth{0pc}
\tablecaption{
Best fit on kinematics with different datasets or different fitting parameters
\label{tab:vdisp_compare}
}
\tablehead{
\colhead{Fitting} & \colhead{\# of stars}  & \colhead{$v_\mathrm{hel}$} & \colhead{$dv/d\Lambda$ or $dv/d\chi$} & \colhead{$\sigma_v$} & \colhead{$\theta_\chi$}\\
 &  & $(\kms)$ & $(\kms\mathrm{deg}^{-1})$ & $(\kms)$ & $(\degr)$
 }
\startdata  	
3-parameter, tails only (default)      &  21     &  \vbulk           &\vgrad                  &  \vdisp                   &   -- \\
4-parameter, tails only                &  21     &  $-101.2\pm0.5$   & $-8.2^{+0.4}_{-0.6}$   &  $0.9^{+0.6}_{-0.5}$      &   $81\pm14$   \\
3-parameter, tails+core                &  30     &  $-101.3\pm0.3$   & $-7.9\pm0.4$           &  $0.6\pm0.4$          &    -- \\
4-parameter, tails+core                 &  30     &  $-101.3\pm0.3$   & $-8.1^{+0.4}_{-0.5}$   &  $0.6\pm0.4$          &   $80\pm12$ \\
3-parameter, core~\citep{tuc3}     &  26     &  $-102.2\pm0.4$   & $-6.0 \pm 3.9$           &  $<1.3$                   &   --   \\
4-parameter, core~\citep{tuc3}     &  26     &  $-102.2\pm0.4$   & $-6.7 \pm 6.1$           &  $<1.3$                   &  $103^{+43}_{-68}$ \\
11-parameter membership, tail+core     &  552    &  $-101.4\pm0.5$  & $-8.4_{-0.5}^{+0.4} $ &  $0.8 \pm 0.4$      &    $86\pm13$ \\[-0.6em]
\enddata
\tablecomments{
All values reported here (and in this paper) are from the 50th percentile of the posterior probability distributions. The uncertainties are from the 16th and 84th percentiles of the posterior probability distributions. For upper limit, 95\% confidence level is used.
}
\end{deluxetable*}
\clearpage

\begin{deluxetable*}{c r r c c c c c r}
\tabletypesize{\scriptsize}
\tablecaption{Potential members in the Tuc~III stream.
\label{tab:tuc3_spec_pot}
}
\tablehead{ID & $\alpha_{2000}$ & $\delta_{2000}$  & $g$ & $r$ & Cat & Inst & S/N & \multicolumn{1}{c}{$v$} \\ 
 & (deg) & (deg) & (mag) & (mag) &  &   &  & \multicolumn{1}{c}{(\kms)} }

\startdata
DES\,J233957.91$-$593718.3 & 354.99128 & -59.62174 & 19.392 & 18.950 & DR1 & AAT &  5.1 & -126.85 \\
DES\,J234209.92$-$594724.9 & 355.54135 & -59.79024 & 19.806 & 19.398 & DR1 & AAT &  2.4 & -129.83 \\
DES\,J234258.94$-$591925.4 & 355.74557 & -59.32372 & 19.784 & 19.363 & DR1 & AAT &  6.0 & -112.15 \\
DES\,J234310.16$-$595556.6 & 355.79232 & -59.93240 & 20.956 & 20.758 & DR1 & AAT &  3.1 &  -82.66 \\
DES\,J234311.08$-$590221.5 & 355.79615 & -59.03932 & 19.330 & 18.851 & DR1 & AAT &  2.2 &  -73.82 \\
DES\,J234325.17$-$594435.1 & 355.85489 & -59.74308 & 18.763 & 18.320 & DR1 & AAT &  4.5 &  -82.15 \\
DES\,J234354.41$-$600605.7 & 355.97671 & -60.10158 & 20.548 & 20.333 & DR1 & AAT &  1.7 & -108.10 \\
DES\,J234446.62$-$601636.3 & 356.19427 & -60.27674 & 19.777 & 19.363 & DR1 & AAT &  5.6 & -113.56 \\
DES\,J234502.83$-$594108.5 & 356.26180 & -59.68569 & 20.696 & 20.437 & DR1 & AAT &  2.1 &  -96.39 \\
DES\,J234539.31$-$600534.8 & 356.41378 & -60.09300 & 20.413 & 20.165 & DR1 & AAT &  3.3 &  -85.55 \\
DES\,J234653.40$-$600737.8 & 356.72251 & -60.12717 & 19.906 & 19.502 & DR1 & AAT &  5.3 & -114.54 \\
DES\,J234703.67$-$592837.7 & 356.76528 & -59.47714 & 19.933 & 19.527 & DR1 & AAT &  4.7 & -135.28 \\
DES\,J234722.75$-$601040.8 & 356.84479 & -60.17799 & 20.271 & 19.968 & DR1 & AAT &  2.0 & -100.87 \\
DES\,J234754.05$-$593615.2 & 356.97523 & -59.60422 & 20.103 & 19.714 & DR1 & AAT &  4.2 & -117.72 \\
DES\,J234801.37$-$591054.6 & 357.00572 & -59.18182 & 20.386 & 20.146 & DR1 & AAT &  3.1 & -117.09 \\
DES\,J234822.42$-$600631.7 & 357.09343 & -60.10880 & 20.910 & 20.687 & DR1 & AAT &  6.5 & -103.71 \\
DES\,J234854.73$-$593421.6 & 357.22804 & -59.57266 & 20.534 & 20.302 & DR1 & AAT &  1.9 & -132.98 \\
DES\,J234907.03$-$595451.4 & 357.27929 & -59.91428 & 20.790 & 20.569 & DR1 & AAT &  3.7 &  -90.85 \\
DES\,J234916.70$-$595028.3 & 357.31959 & -59.84118 & 20.457 & 20.231 & DR1 & AAT &  3.8 & -125.84 \\
DES\,J234934.85$-$594127.9 & 357.39519 & -59.69107 & 20.718 & 20.501 & DR1 & AAT &  1.9 & -130.20 \\
DES\,J234955.16$-$593738.0 & 357.47985 & -59.62722 & 20.374 & 20.128 & DR1 & AAT &  4.0 & -112.21 \\
DES\,J235011.85$-$592433.7 & 357.54936 & -59.40936 & 20.308 & 20.025 & DR1 & AAT &  3.3 & -137.01 \\
DES\,J235105.24$-$594437.3 & 357.77182 & -59.74369 & 20.372 & 20.098 & DR1 & AAT &  2.8 & -111.33 \\
DES\,J235134.95$-$594124.6 & 357.89564 & -59.69016 & 19.846 & 19.411 & DR1 & AAT &  3.9 & -110.51 \\
DES\,J235151.98$-$594056.9 & 357.96659 & -59.68247 & 19.806 & 19.377 & DR1 & AAT &  6.2 & -111.75 \\
DES\,J235209.62$-$590450.5 & 358.04007 & -59.08070 & 19.844 & 19.411 & DR1 & AAT &  5.8 & -111.49 \\
DES\,J235258.81$-$594111.3 & 358.24505 & -59.68646 & 20.553 & 20.337 & DR1 & AAT &  1.6 & -123.43 \\
DES\,J235318.39$-$593243.8 & 358.32662 & -59.54549 & 20.422 & 20.179 & DR1 & IMACS &  4.5 & -100.10 \\
DES\,J235328.33$-$593731.2 & 358.36804 & -59.62533 & 21.523 & 21.271 & DR1 & IMACS &  2.3 & -107.67 \\
DES\,J235341.40$-$592600.9 & 358.42251 & -59.43358 & 19.796 & 19.329 & DR1 & AAT &  3.6 &  -93.32 \\
DES\,J235400.65$-$593255.5 & 358.50273 & -59.54876 & 20.414 & 20.195 & DR1 & IMACS &  5.5 &  -99.64 \\
DES\,J235408.52$-$594422.1 & 358.53551 & -59.73947 & 20.282 & 19.984 & DR1 & AAT &  2.8 &  -83.00 \\
DES\,J235425.68$-$595943.3 & 358.60701 & -59.99537 & 20.313 & 20.002 & DR1 & AAT &  1.9 &  -93.49 \\
DES\,J235425.87$-$594042.6 & 358.60780 & -59.67849 & 20.494 & 20.269 & DR1 & IMACS &  5.7 & -107.18 \\
DES\,J235439.51$-$594118.7 & 358.66463 & -59.68854 & 19.997 & 19.611 & DR1 & AAT &  4.3 &  -83.00 \\
DES\,J235544.46$-$591918.1 & 358.93527 & -59.32169 & 20.427 & 20.185 & DR1 & AAT &  4.3 & -120.70 \\
DES\,J235549.61$-$592446.4 & 358.95673 & -59.41289 & 20.274 & 19.894 & DR1 & AAT &  3.1 & -135.94 \\
DES\,J235615.37$-$595231.4 & 359.06405 & -59.87540 & 20.314 & 20.035 & DR1 & AAT &  2.3 &  -71.16 \\
DES\,J235619.28$-$592219.0 & 359.08034 & -59.37195 & 20.788 & 20.560 & DR1 & AAT &  2.0 &  -93.64 \\
DES\,J235634.89$-$593001.2 & 359.14537 & -59.50032 & 21.268 & 21.056 & DR1 & AAT &  2.6 & -108.97 \\
DES\,J235712.70$-$600747.5 & 359.30291 & -60.12986 & 20.004 & 19.608 & DR1 & AAT &  3.1 & -116.45 \\
DES\,J235901.96$-$592204.5 & 359.75815 & -59.36790 & 20.071 & 19.681 & DR1 & AAT &  2.1 &  -87.20 \\
DES\,J000101.57$-$593029.7 &   0.25654 & -59.50824 & 19.631 & 19.184 & DR1 & AAT &  4.5 &  -96.29 \\
DES\,J000105.45$-$592448.8 &   0.27272 & -59.41355 & 19.726 & 19.265 & DR1 & AAT &  4.3 & -100.99 \\
DES\,J000133.88$-$595528.8 &   0.39116 & -59.92467 & 20.390 & 20.171 & DR1 & AAT &  2.2 & -106.77 \\
DES\,J000336.20$-$595212.8 &   0.90082 & -59.87023 & 20.262 & 19.941 & DR1 & AAT &  1.8 & -100.49 \\
DES\,J000400.30$-$593024.9 &   1.00126 & -59.50691 & 19.645 & 19.201 & DR1 & AAT &  4.5 &  -94.87 \\
DES\,J000505.18$-$593122.1 &   1.27159 & -59.52280 & 20.825 & 20.620 & DR1 & AAT &  2.0 & -117.23 \\
DES\,J000731.43$-$593438.7 &   1.88097 & -59.57741 & 20.819 & 20.592 & DR1 & AAT &  5.4 & -104.87 \\
DES\,J000826.30$-$593212.2 &   2.10958 & -59.53672 & 19.218 & 18.742 & DR1 & AAT &  6.7 &  -93.81 \\
DES\,J000841.53$-$591007.3 &   2.17302 & -59.16870 & 19.460 & 18.995 & DR1 & AAT &  2.7 &  -86.19 \\
DES\,J000902.34$-$594247.1 &   2.25975 & -59.71310 & 19.828 & 19.409 & DR1 & AAT &  4.7 &  -71.10 \\
\enddata
\tablecomments{
Stars that are observed with AAT or IMACS and have measured RVs in the range of $-140~\kms < v < -70~\kms$, but the S/N of the spectra are too low to obtain robust velocity measurements, and therefore these stars are not included in the analysis. RVs should be used with caution. See details in Appendix~\ref{sec:moremember}.
}
\end{deluxetable*}

\end{document}